\documentclass{elsart}
\usepackage{amsmath}
\usepackage{amssymb}
\usepackage{graphicx}
\usepackage{color}
\usepackage{xy}
\xyoption{all}

\newenvironment{Theorem}{\begin{thm}}{\end{thm}}
\newenvironment{Lemma}{\begin{lem}}{\end{lem}}
\newenvironment{Corollary}{\begin{cor}}{\end{cor}}
\newenvironment{Remark}{\begin{rem}\rm}{\qed\end{rem}}

\newenvironment{Example}{\begin{exmp}\rm}{\qed\end{exmp}}
\newenvironment{Definition}{\begin{defn}\rm}{\end{defn}}
\newenvironment{Proof}{\begin{pf}}{\qed\end{pf}}

\renewcommand{\emptyset}{\varnothing}

\begin{document}

\begin{frontmatter}

\title{Reducibility of Gene Patterns in Ciliates using the Breakpoint Graph\thanksref{NWO}}

\thanks[NWO]{Supported by the Netherlands Organization for
Scientific Research (NWO) project 635.100.006 `VIEWS'.}

\author[Leiden]{Robert Brijder\corauthref{cor}},
\corauth[cor]{Corresponding author.}
\ead{rbrijder@liacs.nl}
\author[Leiden]{Hendrik Jan Hoogeboom}, and
\author[Leiden,Boulder]{Grzegorz Rozenberg}

\address[Leiden]{Leiden Institute of Advanced Computer Science,
Universiteit Leiden,\\ Leiden, The Netherlands}

\address[Boulder]{Department of Computer Science,
University of Colorado,\\ Boulder, Colorado, USA}

\begin{abstract}
Gene assembly in ciliates is one of the most involved DNA
processings going on in any organism. This process transforms one
nucleus (the micronucleus) into another functionally different
nucleus (the macronucleus).
We continue the development of the theoretical models of gene
assembly, and in particular we demonstrate the use of the concept of
the breakpoint graph, known from another branch of DNA
transformation research.
More specifically: (1) we characterize the \emph{intermediate} gene
patterns that can occur during the transformation of a \emph{given}
micronuclear gene pattern to its macronuclear form; (2) we determine
the number of applications of the loop recombination operation (the
most basic of the three molecular operations that accomplish gene
assembly) needed in this transformation; (3) we generalize previous
results (and give elegant alternatives for some proofs) concerning
characterizations of the micronuclear gene patterns that can be
assembled using a specific subset of the three molecular operations.
\end{abstract}

\end{frontmatter}

\section{Introduction}
Ciliates are single cell organisms that have two functionally
different nuclei, one called micronucleus and the other called
macronucleus (both of which can occur in various multiplicities). At
some stage in sexual reproduction a micronucleus is transformed into
a macronucleus in a process called gene assembly. This is the most
involved DNA processing in living organisms known today. The reason
that gene assembly is so involved is that the genome of the
micronucleus may be dramatically different from the genome of the
macronucleus --- this is particularly true in the stichotrichs group
of ciliates, which we consider in this paper. The investigation of
gene assembly turns out to be very exciting from both biological and
computational points of view.

Another research area concerned with transformations of DNA is
\emph{sorting by reversal}, see, e.g.,
\cite{SetubalMeidanisBook,PevznerBook,DBLP:conf/recomb/BergeronMS05}.
Two different species can have several contiguous segments in their
genome that are very similar, although there relative order (and
orientation) may differ in both genomes. In the theory of sorting by
reversal one tries to determine the number of operations needed to
reorder such a series of genomic `blocks' from one species into that
of another. An essential tool is the \emph{breakpoint graph} (or
reality and desire diagram) which is used to capture both the
present situation, the genome of the first species, and the desired
situation, the genome of the second species.

Motivated by the breakpoint graph, we introduce the notion of
\emph{reduction graph} into the theory of gene assembly. The
intuition of `reality and desire' remains in place, but the
technical details are different. Instead of one operation, the
reversal, we have three operations. Furthermore, these operations
are irreversible and can only be applied on special positions in the
string, called \emph{pointers}. Also, instead of two different
species, we deal with two different nuclei
--- the reality is a gene in its micronuclear form, and desire
is the same gene but in its macronuclear form. Surprisingly, where
the breakpoint graph in the theory of sorting by reversal is mostly
useful to determine the number of needed operations, the reduction
graph has different uses in the theory of gene assembly, providing
valuable insights into the gene assembly process. Adapted from the
theory of sorting by reversal, and applied to the theory of gene
assembly in ciliates, we hope the reduction graph can serve as a
`missing link' to connect the two fields.

For example, the reduction graph allows for a direct
characterization of the \emph{intermediate} strings that may be
constructed during the transformation of a given gene from its
micronuclear form to its macronuclear form
(Theorem~\ref{main_theorem}). Also, it makes the number of loop
recombination operations (see Figure~\ref{ld_op_fig} below) needed
in this transformation quite explicit as the number of cyclic
(connected) components in the reduction graph
(Theorem~\ref{th_cyclic_components2}).

Each micronuclear form of a gene defines a sequence of (oriented)
segments, the boundaries of which define the pointers where splicing
takes place. In abstract representation, the gene defines a
so-called \emph{realistic} string in which every pointer is denoted
by a single symbol. Each pointer occurs twice (up to inversion) in
that string. Not every string in which each symbol has two
occurrences (up to inversion) can be obtained as the representation
of a micronuclear gene. Our results are obtained in the larger
context, i.e., they are not only valid for realistic strings, but
for \emph{legal} strings in general.

The paper is organized as follows. In Section~\ref{paragr_gene
assembly} we briefly discuss the basics of gene assembly in
ciliates, and describe three molecular operations stipulated to
accomplish gene assembly. The reader is referred to monograph
\cite{GeneAssemblyBook} for more background information. In
Section~\ref{paragr_basic_notions} we recall some basic notions and
notation concerning strings and graphs, and then in
Section~\ref{paragr_sprs} we recall the string pointer reduction
system, which is a formal model of gene assembly. This model is used
throughout the rest of this paper. In
Section~\ref{paragr_pointer_rem} we introduce the operation of
pointer removal, which forms a useful formal tool in this paper.
Then in Sections~\ref{paragr_red_graph} and \ref{paragr_rgr} we
introduce our main construct, the reduction graph, and discuss the
transformations of it that correspond to the three molecular
operations. In Section~\ref{paragr_main_result} we provide a
characterization of intermediate forms of a gene resulting from its
assembly to the macronuclear form --- then, in
Section~\ref{paragr_cyclic_comp} we determine the number of loop
recombination operations required in this assembly. As an
application of this last result, in
Section~\ref{paragr_successfulness} we generalize some well-known
results from \cite{SuccessfulnessChar_Original} (and Chapter~13 in
\cite{GeneAssemblyBook}) as well as give elegant alternatives for
these proofs. A conference edition of this paper, containing
selected results without proofs, was presented at
CompLife~\cite{DBLP:conf/complife/BrijderHR05}.

\section{Background: Gene Assembly in Ciliates}
\label{paragr_gene assembly} This section discusses the biological
origin for the string pointer reduction system, the formal model we
discuss in Section~\ref{paragr_sprs} and use throughout this paper.
Let us recall that the \emph{inversion} of a double stranded DNA
sequence $M$, denoted by $\bar M$, is the point rotation of $M$ by
180 degrees. For example, if $M =
\begin{array}{c} GACGT \\ CTGCA \end{array}$, then $\bar M =
\begin{array}{c} ACGTC \\ TGCAG \end{array}$.

\begin{figure}
\begin{center}
\input{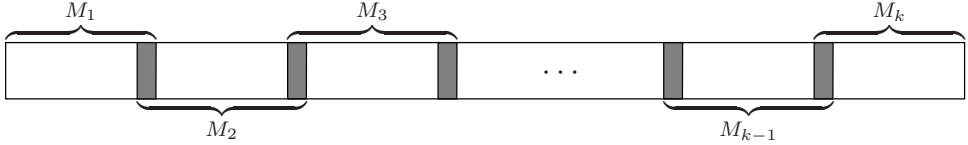} \caption{The MAC form of genes.}\label{MAC_fig}
\end{center}
\end{figure}

\begin{figure}
\begin{center}
\input{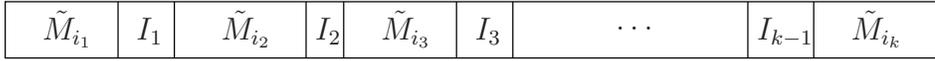} \caption{The MIC form of genes.}\label{MIC_fig}
\end{center}
\end{figure}

Ciliates are unicellular organisms (eukaryotes) that have two kinds
of functionally different nuclei: the micronucleus (MIC) and the
macronucleus (MAC). All the genes occur in both MIC and MAC, but in
very different forms. For a given individual gene (in given species)
the relationship between its MAC and MIC form can be described as
follows.

The MAC form $G$ of a given gene can be represented as the sequence
$M_{1},M_{2},\ldots,M_{k}$ of overlapping segments (called MDSs)
which form $G$ in the way shown in Figure~\ref{MAC_fig} (where the
overlaps are given by the shaded areas). The MIC form $g$ of the
same gene is formed by a specific permutation
$M_{i_{1}},\ldots,M_{i_{k}}$ of $M_{1},\ldots,M_{k}$ in the way
shown in Figure~\ref{MIC_fig}, where $I_{1},I_{2},\ldots,I_{k-1}$
are segments of DNA (called IESs) inserted in-between segments
$\tilde{M}_{i_{1}},\ldots,\tilde{M}_{i_{k}}$ with each $\tilde{M}_i$
equal to either $M_i$ or $\bar{M}_i$ (the inversion of $M_i$). As
clear from Figure~\ref{MAC_fig}, each MDS $M_{i}$ except for $M_{1}$
and $M_{k}$ (the first and the last one) begins with the overlap
with $M_{i-1}$ and ends with the overlap with $M_{i+1}$ --- these
overlap areas are called pointers; the former is the incoming
pointer of $M_{i}$ denoted by $p_{i}$, and the latter is the
outgoing pointer of $M_{i}$ denoted by $p_{i+1}$. Then $M_{1}$ has
only the outgoing pointer $p_{2}$, and $M_{k}$ has only the incoming
pointer $p_{k}$.

The MAC is the (standard eukaryotic) `household' nucleus that
provides RNA transcripts for the expression of proteins --- hence
MAC genes are functional expressible genes. On the other hand the
MIC is a dormant nucleus where no production of RNA transcripts
occurs. As a matter of fact MIC becomes active only during sexual
reproduction. Within a part of sexual reproduction in a process
called \emph{gene assembly}, MIC genes are transformed into MAC
genes (as MIC is transformed into MAC). In this transformation the
IESs from the MIC gene $g$ (see Figure~\ref{MIC_fig}) must be
excised and the MDSs must be spliced (overlapping on pointers) in
their order $M_{1},\ldots,M_{k}$ to form the MAC gene $G$ (see
Figure~\ref{MAC_fig}).

The gene assembly process is accomplished through the following
three molecular operations, which through iterative applications
beginning with the MIC form $g$ of a gene, and going through
intermediate forms, lead to the formation of the MAC form $G$ of the
gene.

\begin{figure}
\begin{center}
\input{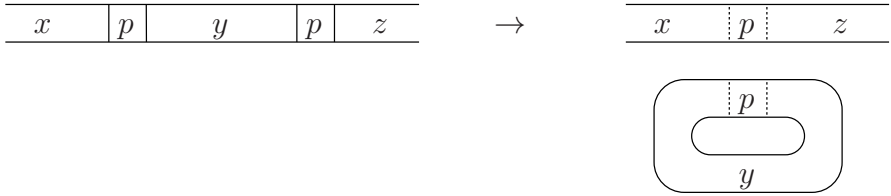}
\caption{The loop recombination operation.}\label{ld_op_fig}
\end{center}
\end{figure}

\begin{figure}
\begin{center}
\input{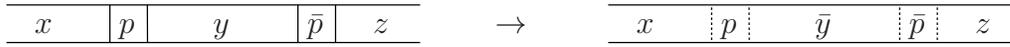}
\caption{The hairpin recombination operation.}\label{hi_op_fig}
\end{center}
\end{figure}

\begin{figure}
\begin{center}
\input{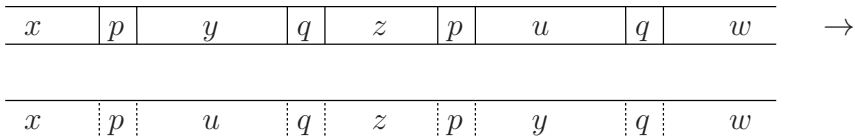}
\caption{The double-loop recombination
operation.}\label{dlad_op_fig}
\end{center}
\end{figure}

\begin{description}
\item[Loop recombination] The effect of the loop recombination operation is
illustrated in Figure~\ref{ld_op_fig}. The operation is applicable
to a gene pattern (i.e., MIC or an intermediate form of a gene)
which has two identical pointers $p$, $p$ separated by a single IES
$y$. The application of this operation results in the excision from
the DNA molecule of a circular molecule consisting of $y$ (and a
copy of the involved pointer) only.
\item[Hairpin recombination] The effect of the hairpin recombination operation
is illustrated in Figure~\ref{hi_op_fig}. The operation is
applicable to a gene pattern containing a pair of pointers $p$,
$\bar p$ in which one pointer is an inversion of the other. The
application of this operation results in the inversion of the DNA
molecule segment that is contained between the mentioned pair of
pointers.
\item[Double-loop recombination] The effect of the double-loop recombination
operation is illustrated in Figure~\ref{dlad_op_fig}. The operation
is applicable to a gene pattern containing two identical pairs of
pointers for which the segment of the molecule between the first
pair of pointers overlaps with the segment of the molecule between
the second pair of pointers. The application of this operation
results in interchanging the segment of the molecule between the
first two (of the four) pointers in the gene pattern and the segment
of the molecule between the last two (of the four) pointers in the
gene pattern.
\end{description}
For a given MIC gene $g$, a sequence of (applications of) these
molecular operations is \emph{successful} if it transforms $g$ into
its MAC form $G$. The gluing of MDS $M_j$ with MDS $M_{j+1}$ on the
common pointer $p_{j+1}$ results in a composite MDS. This means that
after gluing, the outgoing pointer of $M_j$ and the incoming pointer
of $M_{j+1}$ are not pointers anymore, because pointers are always
positioned on the boundary of MDSs (hence they are adjacent to
IESs). Therefore, the molecular operations can be seen as operations
that remove pointers. This is an important property of gene assembly
which is crucial in the formal models of the gene assembly process
(see \cite{GeneAssemblyBook}).

\section{Basic Notions and Notation} \label{paragr_basic_notions}
In this section we recall some basic notions concerning functions,
strings, and graphs. We do this mainly to set up the basic notation
and terminology for this paper.

The empty set will be denoted by $\emptyset$. The composition of
functions $f: X \rightarrow Y$ and $g: Y \rightarrow Z$ is the
function $g f: X \rightarrow Z$ such that $(g f) (x) = g(f(x))$ for
every $x \in X$. The restriction of $f$ to a subset $A$ of $X$ is
denoted by $f|A$.

We will use $\lambda$ to denote the empty string. For strings $u$
and $v$, we say that $v$ is a \emph{substring of $u$} if $u = w_1 v
w_2$, for some strings $w_1$, $w_2$; we also say that $v$
\emph{occurs in $u$}. For a string $x = x_1 x_2 \ldots x_n$ over
$\Sigma$ with $x_1, x_2, \ldots, x_n \in \Sigma$, we say that
substrings $x_{i_1} \cdots x_{j_1}$ and $x_{i_2} \cdots x_{j_2}$ of
$x$ \emph{overlap in $x$} if $i_1 < i_2 < j_1 < j_2$ or $i_2 < i_1 <
j_2 < j_1$.

For alphabets $\Sigma$ and $\Delta$, a \emph{homomorphism} is a
function $\varphi: \Sigma^* \rightarrow \Delta^*$ such that
$\varphi(xy) = \varphi(x)\varphi(y)$ and for all $x,y \in \Sigma^*$.
Let $\varphi: \Sigma^* \rightarrow \Delta^*$ be a homomorphism. If
there is a $\Gamma \subseteq \Sigma$ such that
$$
\varphi(x) = \begin{cases} x & x \not\in \Gamma \\ \lambda & x \in
\Gamma \end{cases},
$$
then $\varphi$ is denoted by $erase_{\Gamma}$.

We move now to graphs. A \emph{labelled graph} is a 4-tuple $G =
(V,E,f,\Psi)$, where $V$ is a finite set, $\Psi$ is an alphabet, $E$
is a finite subset of $V \times \Psi^* \times V$, and $f: D
\rightarrow \Gamma$, for some $D \subseteq V$ and some alphabet
$\Gamma$, is a partial function on $V$. The elements of $V$ are
called \emph{vertices}, and the elements of $E$ are called
\emph{edges}. Function $f$ is the \emph{vertex labelling function},
the elements of $\Gamma$ are the \emph{vertex labels}, and the
elements of $\Psi^*$ are the \emph{edge labels}.

For $e = (x,u,y) \in V \times \Psi^* \times V$, $x$ is called the
\emph{initial vertex} of $e$, denoted by $\iota(e)$, $y$ is called
the \emph{terminal vertex} of $e$, denoted by $\tau(e)$, and $u$ is
called the \emph{label} of $e$, denoted by $\ell(e)$. Labelled graph
$G' = (V',E',f|V',\Psi)$ is an \emph{induced subgraph of $G$} if $V'
\subseteq V$ and $E' = E \cap (V' \times \Psi^* \times V')$. We also
say that $G'$ is the \emph{subgraph of $G$ induced by $V'$}.

A \emph{walk in $G$} is a string $\pi = e_1 e_2 \cdots e_n$ over $E$
with $n \geq 1$ such that $\tau(e_i) = \iota(e_{i+1})$ for $1 \leq i
< n$. The \emph{label of $\pi$} is the string $\ell(\pi) = \ell(e_1)
\ell(e_2) \cdots \ell(e_n)$. Vertex $\iota(e_1)$ is called the
\emph{initial vertex of $\pi$}, denoted by $\iota(\pi)$, vertex
$\tau(e_n)$ is called the \emph{terminal vertex of $\pi$}, denoted
by $\tau(\pi)$ and we say that $\pi$ is a \emph{walk between
$\iota(\pi)$ and $\tau(\pi)$} (or that $\pi$ is a \emph{walk from
$\iota(\pi)$ to $\tau(\pi)$}). We say that $G$ is \emph{weakly
connected} if for every two vertices $v_1$ and $v_2$ of $G$ with
$v_2 \not= v_1$, there is string $e_1 e_2 \cdots e_n$ over $E \cup
\{ (\tau(e),\ell(e),\iota(e)) \mid e \in E\}$ with $n \geq 1$,
$\iota(e_1) = v_1$, $\tau(e_n) = v_2$, and $\tau(e_i) =
\iota(e_{i+1})$ for $1 \leq i < n$. A subgraph $H$ of $G$ induced by
$V_H \subseteq V$ is a \emph{component of $G$} if $H$ is weakly
connected, and for every edge $e \in E$ either $\iota(e), \tau(e)
\in V_H$ or $\iota(e), \tau(e) \in V \backslash V_H$.

The isomorphism between two labelled graphs is defined in the usual
way. Two labelled graphs $G = (V,E,f,\Psi)$ and $G' =
(V',E',f',\Psi)$ are \emph{isomorphic}, denoted by $G \approx G'$,
if there is a bijection $\alpha: V \rightarrow V'$ such that $f(v) =
f'(\alpha(v))$ for all $v \in V$, and
$$
(x,u,y) \in E \mbox{ iff } (\alpha(x),u,\alpha(y)) \in E',
$$
for all $x,y \in V$ and $u \in \Psi^*$. The bijection $\alpha$ is
then called an \emph{isomorphism from $G$ to $G'$}.

In this paper we will consider walks in labelled graphs that often
originate in a fixed source vertex and will end in a fixed target
vertex. Therefore, we need the following notion.

A \emph{two-ended graph} is a 6-tuple $G = (V,E,f,\Psi,s,t)$, where
$(V,E,f,\Psi)$ is a labelled graph, $f$ is a function on $V
\backslash \{s,t\}$ and $s,t \in V$ where $s \not= t$. Vertex $s$ is
called the \emph{source vertex} of $G$ and vertex $t$ is called the
\emph{target vertex} of $G$. The basic notions and notation for
labelled graphs carry over to two-ended graphs. However, for the
notion of isomorphism, care must be taken that the two ends are
preserved. Thus, if $G$ and $G'$ are two-ended graphs, and $\alpha$
is a isomorphism from $G$ to $G'$, then $\alpha(s) = s'$ and
$\alpha(t) = t'$, where $s$ ($s'$, resp.) is the source vertex of
$G$ ($G'$, resp.) and $t$ ($t'$, resp.) is the target vertex of $G$
($G'$, resp.).

\section{The String Pointer Reduction System}
\label{paragr_sprs} In this paper we consider the string pointer
reduction system, which we will recall now (see also
\cite{Equiv_String_Graph_1} and Chapter~9 in
\cite{GeneAssemblyBook}).

\newcommand{\pset}[1]{{\mathbf #1}}
We fix $\kappa \geq 2$, and define the alphabet $\Delta =
\{2,3,\ldots,\kappa\}$. For $D \subseteq \Delta$, we define $\bar D
= \{ \bar a \mid a \in D \}$ and $\Pi_D = D \cup \bar D$; also $\Pi
= \Pi_{\Delta}$. We will use the alphabet $\Pi$ to formally denote
the pointers --- the intuition is that the pointer $p_i$ will be
denoted by either $i$ or $\bar i$. Accordingly, elements of $\Pi$
will also be called \emph{pointers}.

We use the `bar operator' to move from $\Delta$ to $\bar \Delta$ and
back from $\bar \Delta$ to $\Delta$. Hence, for $p \in \Pi$, $\bar
{\bar {p}} = p$. For a string $u = x_1 x_2 \cdots x_n$ with $x_i \in
\Pi$, the \emph{inverse} of $u$ is the string $\bar u = \bar x_n
\bar x_{n-1} \cdots \bar x_1$. For $p \in \Pi$, we define $\pset{p}
=
\begin{cases} p & \mbox{if } p \in \Delta \\ \bar{p} & \mbox{if }
p \in \bar{\Delta}
\end{cases}$, i.e., $\pset{p}$ is the `unbarred' variant of $p$. The
\emph{domain} of a string $v \in \Pi^*$ is $dom(v) = \{ \pset{p}
\mid \mbox{$p$ occurs in $v$} \}$. A \emph{legal string} is a string
$u \in \Pi^*$ such that for each $p \in \Pi$ that occurs in $u$, $u$
contains exactly two occurrences from $\{p,\bar p\}$.

We define the alphabet $\Theta_{\kappa} = \{ M_i, \bar{M}_i \mid 1
\leq i \leq \kappa \}$ --- these symbols denote the MDSs and their
inversions. With each string over $\Theta_{\kappa}$, we associate a
unique string over $\Pi$ through the homomorphism $\pi_{\kappa}:
\Theta^*_{\kappa} \rightarrow \Pi^*$ defined by:
$$
\pi_{\kappa}(M_1) = 2, \quad \pi_{\kappa}(M_{\kappa}) = \kappa,
\quad \pi_{\kappa}(M_i) = i(i+1) \quad \mbox{for } 1 < i < \kappa,
$$
and $\pi_{\kappa}(\bar M_j) = \overline{\pi_{\kappa}(M_j)}$ for $1
\leq j \leq \kappa$. A permutation of the string $M_1 M_2 \cdots
M_{\kappa}$, with possibly some of its elements inverted, is called
a \emph{micronuclear pattern} since it can describe the MIC form of
a gene. String $u$ is \emph{realistic} if there is a micronuclear
pattern $\delta$ such that $u = \pi_{\kappa}(\delta)$.

\begin{Example} \label{ex_actin_protein}
The MIC form of the gene that encodes the actin protein in the
stichotrich \emph{Sterkiella nova} is described by micronuclear
pattern
$$
\delta = M_3 M_4 M_6 M_5 M_7 M_9 \bar{M}_2 M_1 M_8
$$
(see \cite{Biology_of_Ciliates,GeneAssemblyBook}). The associated
realistic string is $\pi_{9}(\delta) = 3 4 4 5 6 7 5 6 7 8 9 \bar 3
\bar 2 2 8 9$.
\end{Example}

Note that every realistic string is legal, but a legal string need
not be realistic. For example, a realistic string cannot have `gaps'
(missing pointers): thus $2244$ is not realistic while it is legal.
It is also easy to produce examples of legal strings which do not
have gaps but still are not realistic --- $3322$ is such an example.
For a pointer $p$ and a legal string $u$, if both $p$ and $\bar p$
occur in $u$ then we say that both $p$ and $\bar p$ are
\emph{positive} in $u$; if on the other hand only $p$ or only $\bar
p$ occurs in $u$, then both $p$ and $\bar p$ are \emph{negative} in
$u$. So, every pointer occurring in a legal string is either
positive or negative in it. A nonempty legal string with no proper
nonempty legal substrings is called \emph{elementary}. For example,
the legal string $234324$ is elementary, while the legal string
$234342$ is not (because $3434$ is a proper legal substring).

\begin{Definition}
Let $u = x_1 x_2 \cdots x_n$ be a legal string with $x_i \in \Pi$
for $1 \leq i \leq n$. For a pointer $p \in \Pi$ such that
$\{x_i,x_j\} \subseteq \{p,\bar p\}$ and $1 \leq i < j \leq n$, the
\emph{p-interval} of $u$ is the substring $x_i x_{i+1} \cdots x_j$.
Two distinct pointers $p,q \in \Pi$ \emph{overlap} in $u$ if the
$p$-interval of $u$ overlaps with the $q$-interval of $u$.
\end{Definition}

The string pointer reduction system consists of three types of
reduction rules operating on legal strings. For all $p,q \in \Pi$
with $\pset{p} \not = \pset{q}$:
\begin{itemize}
\item
the \emph{string negative rule} for $p$ is defined by ${\bf
snr}_{p}(u_1 p p u_2) = u_1 u_2$, \item
the \emph{string positive rule} for $p$ is defined by ${\bf
spr}_{p}(u_1 p u_2 \bar p u_3) = u_1 \bar u_2 u_3$, \item
the \emph{string double rule} for $p,q$ is defined by ${\bf
sdr}_{p,q}(u_1 p u_2 q u_3 p u_4 q u_5) = u_1 u_4 u_3 u_2 u_5$,
\end{itemize}
where $u_1,u_2,\ldots,u_5$ are arbitrary strings over $\Pi$.

Note that each of these rules is defined only on legal strings that
satisfy the given form. For example, ${\bf snr}_{2}$ is not defined
on legal string $2323$. It is important to realize that for every
non-empty legal string there is at least one reduction rule
applicable. Indeed, every legal string for which no string positive
rule and no string double rule is applicable must have only
nonoverlapping, negative pointers and thus a string negative rule is
applicable.

We also define $Snr = \{ {\bf snr}_p \mid p \in \Pi \}$, $Spr = \{
{\bf spr}_p \mid p \in \Pi \}$ and $Sdr = \{ {\bf sdr}_{p,q} \mid
p,q \in \Pi, \pset{p} \not = \pset{q} \}$ to be the sets containing
all the reduction rules of a specific type.

The string negative rule corresponds to the loop recombination
operation, the string positive rule corresponds to the hairpin
recombination operation, and the string double rule corresponds to
the double-loop recombination operation. Note that the fact (pointed
out at the end of Section~\ref{paragr_gene assembly}) that the
molecular operations remove pointers is explicit in the string
pointer reduction system --- indeed when a string rule for a pointer
$p$ (or pointers $p$ and $q$) is applied, then all occurrences of
$p$ and $\bar p$ (or $p$, $\bar p$, $q$ and $\bar q$) are removed.

\begin{Definition}
The \emph{domain} $dom(\rho)$ of a reduction rule $\rho$ equals the
set of unbarred variants of the pointers the rule is applied to,
i.e., $dom({\bf snr}_p) = dom({\bf spr}_p) = \{\pset{p}\}$ and
$dom({\bf sdr}_{p,q}) = \{\pset{p},\pset{q}\}$ for $p,q \in \Pi$.
For a composition $\varphi = \varphi_1 \ \varphi_2 \ \cdots \
\varphi_n$ of reduction rules $\varphi_1, \varphi_2, \ldots,
\varphi_n$, the \emph{domain} $dom(\varphi)$ is the union of the
domains of its constituents, i.e., $dom(\varphi) = dom(\varphi_1)
\cup dom(\varphi_2) \cup \cdots \cup dom(\varphi_n)$.
\end{Definition}

\begin{Definition}
Let $u$ and $v$ be legal strings and $S \subseteq \{Snr,Spr,Sdr\}$.
Then a composition $\varphi$ of reduction rules from $S$ is called
an \emph{($S$-)reduction of $u$}, if $\varphi$ is applicable to
(defined on) $u$. A \emph{successful reduction $\varphi$ of $u$} is
a reduction of $u$ such that $\varphi(u) = \lambda$. We then also
say that $\varphi$ is \emph{successful for $u$}. We say that $u$ is
\emph{reducible to $v$ in $S$} if there is a $S$-reduction $\varphi$
of $u$ such that $\varphi(u)=v$. We simply say that $u$ is
\emph{reducible to $v$} if $u$ is reducible to $v$ in
$\{Snr,Spr,Sdr\}$. We say that $u$ is \emph{successful in $S$} if
$u$ is reducible to $\lambda$ in $S$.
\end{Definition}
Note that if $\varphi$ is a reduction of $u$, then $dom(\varphi) =
dom(u) \backslash dom(\varphi(u))$. Because (as pointed out already)
for every non-empty legal string there is at least one reduction
rule applicable, we easily obtain Theorem 9.1 in
\cite{GeneAssemblyBook} which states that every legal string is
successful in $\{Snr,Spr,Sdr\}$.

\begin{Example} Let $S = \{Snr,Spr\}$, $u = 3 2 4 5 \bar 4 5 \bar
3 \bar 2$, and $v = \bar 5 4 \bar 5 \bar 4$. Then $u$ is reducible
to $v$ in $S$, because $({\bf snr}_3 \ {\bf spr}_2)(u) = v$. Since
applying $\varphi = {\bf spr}_{\bar 5} \ {\bf spr}_{4} \ {\bf
snr}_{\bar 2} \ {\bf spr}_{3}$ to $u$ yields $\lambda$, $\varphi$ is
successful for $u$. On the other hand, $u = 3232$ is not reducible
to any $v$ in $S$, because none of the rules in $Snr$ and none of
the rules in $Spr$ is applicable for this $u$.
\end{Example}

Referring to the Introduction, in Theorem~\ref{main_theorem} we
present a characterization of the intermediate strings that may be
constructed during the transformation of a given gene from its
micronuclear form to its macronuclear form. Formally, this is a
characterization of reducibility, which allows one to determine for
any given legal strings $u$ and $v$ and $S \subseteq
\{Snr,Spr,Sdr\}$, whether or not $u$ is reducible to $v$ in $S$.
This result can be seen as a generalization of the results from
Chapter~13 in \cite{GeneAssemblyBook}, which provide a
characterization of successfulness for realistic strings, that is,
for the case where $u$ is realistic and $v = \lambda$.

\section{Pointer Removal Operation} \label{paragr_pointer_rem}
Let $\varphi$ be a reduction of a legal string $u$. If we let $u'$
be the legal string obtained from $u$ be deleting all pointers from
$\Pi_{dom(\varphi(u))}$, then it turns out that $\varphi$ is also a
reduction of $u'$. In fact, $\varphi$ is a successful reduction of
$u'$. This is formalized in Theorem~\ref{char1}, and thus it states
a necessary condition for reducibility. In the following sections we
will strengthen Theorem~\ref{char1} to obtain a characterization of
reducibility.

\begin{Definition} For a subset $D \subseteq \Delta$, the $D$-removal operation,
denoted by $rem_D$, is defined by $rem_D = erase_{D \cup \bar{D}}$.
We also refer to $rem_D$ operations, for all $D \subseteq \Delta$,
as \emph{pointer removal operations}.
\end{Definition}

\begin{Example} Let $u = 3 2 4 5 \bar 4 5 \bar 3 \bar 2$ and $D =
\{ 4, 5 \}$. Then $rem_D(u) = 3 2 \bar 3 \bar 2$. Note that $2,3
\not\in D$. Note also that $\varphi = {\bf snr}_3 \ {\bf spr}_2$ is
applicable to both $u$ and $rem_D(u)$, but for $rem_D(u)$, $\varphi$
is also successful.
\end{Example}

The following easy to verify lemma formalizes the essence of the
above example.
\begin{Lemma} \label{ops_appl}
Let $u$ be a legal string and $D \subseteq dom(u)$. Let $\varphi$ be
a composition of reduction rules.
\begin{enumerate}
\item
If $\varphi$ is applicable to $rem_D(u)$ and $\varphi$ does not
contain string negative rules, then $\varphi$ is applicable to $u$.
\item
If $\varphi$ is applicable to $u$ and $dom(\varphi) \subseteq dom(u)
\backslash D$, then $\varphi$ is applicable to $rem_D(u)$.
\item
If $\varphi$ is applicable to both $u$ and $rem_D(u)$, then
$\varphi(rem_D(u)) = rem_D(\varphi(u))$.
\end{enumerate}
\end{Lemma}
Note that the first statement of Lemma~\ref{ops_appl} may not be
true when $\varphi$ is allowed to contain string negative rules. The
obvious reason for this is that two identical occurrences of a
pointer $p$ may end up to be next to each other only if some
pointers in between those occurrences are first removed by $rem_D$.
This is illustrated in the following example.
\begin{Example} \label{ex2_rem} Let $u = 3 2 4 5 \bar 4 5 \bar 3 6 6 \bar 2$,
$v = \bar 5 4 \bar 5 \bar 4 6 6$ and $D = dom(v)$. Then $rem_D(u) =
3 2 \bar 3 \bar 2$. Note that although $\varphi = {\bf snr}_3 \ {\bf
spr}_2$ is a successful reduction of $rem_D(u)$, $\varphi$ is not
applicable to $u$.
\end{Example}
The following theorem is an immediate consequence of the previous
lemma.
\begin{Theorem} \label{char1}
Let $S \subseteq \{Snr,Spr,Sdr\}$. For legal strings $u$ and $v$, if
$u$ is reducible to $v$ in $S$ and $D = dom(v)$, then $rem_D(u)$ is
successful in $S$.
\end{Theorem}
\begin{Proof}
Let $u$ be reducible to $v$ in $S$. Then there is an $S$-reduction
$\varphi$ such that $\varphi(u)=v$. By Lemma~\ref{ops_appl},
$\varphi$ is an $S$-reduction of $rem_D(u)$ and $\varphi(rem_D(u)) =
rem_D(\varphi(u)) = rem_D(v) = \lambda$. Hence, $\varphi$ is a
successful $S$-reduction of $rem_D(u)$.
\end{Proof}
The proof of the above result observes that any reduction of $u$
into $v$ must be a successful reduction of $rem_D(u)$ where $D =
dom(v)$. Referring to Example~\ref{ex2_rem}, we now note that $u$ is
not reducible to $v$, because $rem_D(u)$ has two successful
reductions and neither is applicable to $u$. In fact, there is no
$v'$ with $D = dom(v')$ such that $u$ is reducible to $v'$.

\section{Reduction Graphs}
\label{paragr_red_graph} The main purpose of this section is to
define the notion of reduction graph. A reduction graph represents
some key aspects of reductions from a legal string $u$ to a legal
string $v$: it provides the additional requirements on $u$ and $v$
to make the reverse implication of Theorem~\ref{char1} hold. In
addition, it allows one to easily determine the number of string
negative rules needed to successfully reduce $u$. We will first
define the notion of a 2-edge coloured graph.
\begin{Definition}
A \emph{2-edge coloured graph} is a 7-tuple
$$
G = (V,E_1,E_2,f,\Psi,s,t),
$$
where both $(V,E_1,f,\Psi,s,t)$ and $(V,E_2,f,\Psi,s,t)$ are
two-ended graphs. Note that $E_1$ and $E_2$ are not necessary
disjoint.
\end{Definition}

The terminology and notation for the two-ended graph carries over to
2-edge coloured graphs. However, for the notion of isomorphism, care
must be taken that the two sorts of edges are preserved. Thus, if $G
= (V,E_1,E_2,f,\Psi,s,t)$ and $G' = (V',E_1',E_2',f',\Psi,s',t')$
are two-ended graphs, then it must hold that for any isomorphism
$\alpha$ from $G$ to $G'$,
$$
(x,u,y) \in E_i \mbox{ iff } (\alpha(x),u,\alpha(y)) \in E_i'
$$
for all $x,y \in V$, $u \in \Psi$ and $i \in \{1,2\}$.

We say that edges $e_1$ and $e_2$ have the \emph{same colour} if
either $e_1,e_2 \in E_1$ or $e_1,e_2 \in E_2$, otherwise they have
\emph{different colours}. An \emph{alternating walk} in $G$ is a
walk $\pi = e_1 e_2 \cdots e_n$ in $G$ such that $e_i$ and $e_{i+1}$
have different colours for $1 \leq i < n$. For each edge $e$ with
$\ell(e) \in \Pi^*$, we define $(\tau(e),\overline{l(e)},\iota(e))$,
denoted by $\bar e$, as the \emph{reverse of $e$}.

We are ready now to define the notion of a reduction graph, the main
technical notion of this paper. The reduction graph is a 2-edge
coloured graph and it is defined for a legal string $u$ and a set of
pointers $D \subseteq dom(u)$. The intuition behind it is as
follows.

\begin{figure}
\begin{center}
\setlength{\unitlength}{4144sp}%
\begingroup\makeatletter\ifx\SetFigFont\undefined
\def\x#1#2#3#4#5#6#7\relax{\def\x{#1#2#3#4#5#6}}%
\expandafter\x\fmtname xxxxxx\relax \def\y{splain}%
\ifx\x\y   
\gdef\SetFigFont#1#2#3{%
  \ifnum #1<17\tiny\else \ifnum #1<20\small\else
  \ifnum #1<24\normalsize\else \ifnum #1<29\large\else
  \ifnum #1<34\Large\else \ifnum #1<41\LARGE\else
     \huge\fi\fi\fi\fi\fi\fi
  \csname #3\endcsname}%
\else
\gdef\SetFigFont#1#2#3{\begingroup
  \count@#1\relax \ifnum 25<\count@\count@25\fi
  \def\x{\endgroup\@setsize\SetFigFont{#2pt}}%
  \expandafter\x
    \csname \romannumeral\the\count@ pt\expandafter\endcsname
    \csname @\romannumeral\the\count@ pt\endcsname
  \csname #3\endcsname}%
\fi
\fi\endgroup
\begin{picture}(5086,332)(-67,468)
\thinlines
\put(-55,727){\line( 1, 0){5062}}
\put(563,727){\line( 0,-1){225}}
\put(563,502){\line( 1, 0){169}}
\put(732,502){\line( 0, 1){225}}
\put(732,727){\line(-1, 0){169}}
\put(1182,727){\line( 0,-1){225}}
\put(1182,502){\line( 1, 0){169}}
\put(1351,502){\line( 0, 1){225}}
\put(1351,727){\line(-1, 0){169}}
\put(1912,727){\line( 0,-1){225}}
\put(1912,502){\line( 1, 0){169}}
\put(2081,502){\line( 0, 1){225}}
\put(2081,727){\line(-1, 0){169}}
\put(2644,727){\line( 0,-1){225}}
\put(2644,502){\line( 1, 0){169}}
\put(2813,502){\line( 0, 1){225}}
\put(2813,727){\line(-1, 0){169}}
\put(3376,727){\line( 0,-1){225}}
\put(3376,502){\line( 1, 0){169}}
\put(3545,502){\line( 0, 1){225}}
\put(3545,727){\line(-1, 0){169}}
\put(4162,727){\line( 0,-1){225}}
\put(4162,502){\line( 1, 0){169}}
\put(4331,502){\line( 0, 1){225}}
\put(4331,727){\line(-1, 0){169}}
\put(-55,502){\line( 1, 0){5062}}
\put(4219,558){$4$}
\put(620,558){$2$}
\put(1238,558){$3$}
\put(1970,558){$\bar{2}$}
\put(2701,558){$\bar{4}$}
\put(3432,558){$3$}
\end{picture}
\caption{Part of a genome with
three pointer pairs corresponding to the same
gene.}\label{fig_motivation_red_graph1}
\end{center}
\end{figure}
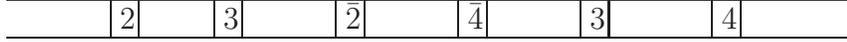
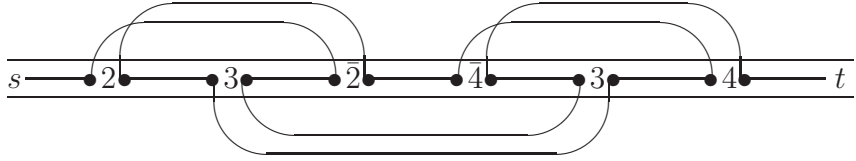
\begin{figure}
\begin{center}
\setlength{\unitlength}{4144sp}%
\begingroup\makeatletter\ifx\SetFigFont\undefined
\def\x#1#2#3#4#5#6#7\relax{\def\x{#1#2#3#4#5#6}}%
\expandafter\x\fmtname xxxxxx\relax \def\y{splain}%
\ifx\x\y   
\gdef\SetFigFont#1#2#3{%
  \ifnum #1<17\tiny\else \ifnum #1<20\small\else
  \ifnum #1<24\normalsize\else \ifnum #1<29\large\else
  \ifnum #1<34\Large\else \ifnum #1<41\LARGE\else
     \huge\fi\fi\fi\fi\fi\fi
  \csname #3\endcsname}%
\else
\gdef\SetFigFont#1#2#3{\begingroup
  \count@#1\relax \ifnum 25<\count@\count@25\fi
  \def\x{\endgroup\@setsize\SetFigFont{#2pt}}%
  \expandafter\x
    \csname \romannumeral\the\count@ pt\expandafter\endcsname
    \csname @\romannumeral\the\count@ pt\endcsname
  \csname #3\endcsname}%
\fi
\fi\endgroup
\begin{picture}(5086,920)(-67,155)
\put(506,558){$2$}
\put(600,558){$\bullet$}
\put(395,558){$\bullet$}
\thinlines
\put(1181,502){\line( 1, 0){169}}
\put(1350,727){\line(-1, 0){169}}
\put(1126,558){$\bullet$}
\put(1331,558){$\bullet$}
\put(1912,502){\line( 1, 0){169}}
\put(2081,727){\line(-1, 0){169}}
\put(2062,558){$\bullet$}
\put(1857,558){$\bullet$}
\put(2644,502){\line( 1, 0){169}}
\put(2813,727){\line(-1, 0){169}}
\put(2588,558){$\bullet$}
\put(2793,558){$\bullet$}
\put(3375,502){\line( 1, 0){169}}
\put(3544,727){\line(-1, 0){169}}
\put(3524,558){$\bullet$}
\put(3319,558){$\bullet$}
\put(4162,502){\line( 1, 0){169}}
\put(4331,727){\line(-1, 0){169}}
\put(4107,558){$\bullet$}
\put(4312,558){$\bullet$}
\put(2364,614){\oval(2362,902)[bl]}
\put(2364,614){\oval(2362,902)[br]}
\put(2363,614){\oval(2024,676)[bl]}
\put(2363,614){\oval(2026,676)[br]}
\put(3573,614){\oval(1518,902)[tr]}
\put(3573,614){\oval(1518,902)[tl]}
\put(3404,614){\oval(1518,676)[tr]}
\put(3404,614){\oval(1518,676)[tl]}
\put(1351,614){\oval(1462,904)[tr]}
\put(1351,614){\oval(1462,904)[tl]}
\put(1182,614){\oval(1462,678)[tr]}
\put(1182,614){\oval(1462,678)[tl]}
\put(4332,614){\line( 1, 0){506}}
\put(3544,614){\line( 1, 0){619}}
\put(2813,614){\line( 1, 0){563}}
\put(2082,614){\line( 1, 0){562}}
\put(1351,614){\line( 1, 0){562}}
\put(619,614){\line( 1, 0){563}}
\put( 57,614){\line( 1, 0){394}}
\put(-55,727){\line( 1, 0){5062}}
\put(-55,502){\line( 1, 0){5062}}
\put(4895,558){$t$}
\put(-54,558){$s$}
\put(4219,558){$4$}
\put(3432,558){$3$}
\put(2701,558){$\bar{4}$}
\put(1970,558){$\bar{2}$}
\put(1238,558){$3$}
\put(450,502){\line( 1, 0){169}}
\put(619,727){\line(-1, 0){169}}
\end{picture}
\caption{The reduction graph corresponding to the underlying
genome.}\label{fig_motivation_red_graph2}
\end{center}
\end{figure}
Figure~\ref{fig_motivation_red_graph1} depicts a part of a genome
with three pointer pairs corresponding to the same gene $g$. The
reduction graph introduces two vertices for each pointer and two
special vertices $s$ and $t$ representing the ends. It connects
adjacent pointers through \emph{reality edges} and connects pointers
corresponding to the same pointer pair through \emph{desire edges}
in a way that reflects how the parts will be glued after a molecular
operation is applied on that pointer. The resulting reduction graph
is depicted in Figure~\ref{fig_motivation_red_graph2}. Thus, every
reality edge corresponds to a certain DNA segment. If such a DNA
segment contains other pointers of $g$, then these pointers form the
label of that reality edge.

By definition a realistic string has a physical interpretation. It
shows the boundaries of the MDSs, and how these should be recombined
(following their orientation). Considering a subset of these
pointers, we still have the physical interpretation, although the
other pointers are hidden in the segments. Technically, however,
removing a subset of the pointers may change a realistic string into
a legal one that is no longer realistic or even realizable (by
renaming pointers we cannot obtain a realistic string). An example
of such a case is given in the introduction of
Section~\ref{paragr_successfulness}. In fact, each legal string has
a physical interpretation with pointers indicating how parts of the
string are to be reconnected, cf.
Figure~\ref{fig_motivation_red_graph2}, where no use is made of any
MDS-IES segmentation. Thus our definition of reduction graph works
for legal strings in general, rather than only for realistic ones.
The intuition of a reduction graph is similar to the intuition
behind a reality and desire diagram (or breakpoint graph) from
\cite{Breakpoint_Graph,PevznerBook}.

Formally, the reduction graph of legal string $u$ with respect to $D
\subseteq dom(u)$ shows how $u$ is reduced to a legal string $v$
with $dom(v) = D$ by any possible reduction $\varphi$. The vertices
of the graph correspond to (two copies of each of) the pointers that
are removed during the reduction (those in $\Pi_{dom(u) \backslash
D}$). As illustrated above, we have two types of edges. The desire
edges are unlabelled and connect the pointer pairs in $\Pi_{dom(u)
\backslash D}$, while reality edges connect the successive pointers
in $\Pi_{dom(u) \backslash D}$ and are labelled by the strings over
$\Pi^*_D$ that are in between these pointers in $u$.

\newcommand{\RGVertL}[1]{I_{#1}}
\newcommand{\RGVertR}[1]{I'_{#1}}
\newcommand{\redgr}{\mathcal{R}}
\begin{Definition}
Let $D \subseteq \Delta$ and let $u$ be a legal string, such that $u
= \delta_0 p_1 \delta_1 p_2 \ldots p_n \delta_n$ where
$\delta_0,\ldots,\delta_n \in \Pi^*_D$ and $p_1,\ldots,p_n \in
\Pi_{dom(u) \backslash D}$. The \emph{reduction graph of $u$ with
respect to $D$}, denoted by $\redgr_{u,D}$, is a 2-edge coloured
graph $(V,E_1,E_2,f,\Pi,s,t)$, where
$$
V = \{\RGVertL{1},\RGVertL{2},\ldots,\RGVertL{n}\} \ \cup \
\{\RGVertR{1},\RGVertR{2},\ldots,\RGVertR{n}\} \ \cup \ \{s,t\},
$$
$$
E_1 = E_{1,r} \ \cup \ E_{1,l}, \mbox{where}
$$
$$
E_{1,r} = \{ e_0, e_1, \ldots, e_n \} \mbox{ with }  e_i =
(\RGVertR{i},\delta_i,\RGVertL{i+1}) \mbox{ for } 1 \leq i \leq n-1,
e_0 = (s,\RGVertL{1}), e_n = (\RGVertR{n},t),
$$
$$
E_{1,l} = \{ \bar e \mid e \in E_{1,r}\},
$$
\begin{eqnarray*}
E_{2} = & \{ (\RGVertR{i},\lambda,\RGVertL{j}),
(\RGVertL{i},\lambda,\RGVertR{j}) \mid
i,j \in \{1,2,\ldots,n\} \mbox{ with } i \not= j \mbox{ and } p_i = p_j \} \ \cup \ \\
& \{ (\RGVertL{i},\lambda,\RGVertL{j}),
(\RGVertR{i},\lambda,\RGVertR{j}) \mid i,j \in \{1,2,\ldots,n\}
\mbox{ and } p_i = \bar{p}_j \}, \mbox{and}
\end{eqnarray*}
$$
\mbox{$f(\RGVertL{i}) = f(\RGVertR{i}) = \pset{p_i}$ for $1 \leq i
\leq n$.}
$$
\mbox{ }
\end{Definition}

The edges of $E_1$ are called the \emph{reality edges}, and the
edges of $E_2$ are called the \emph{desire edges}. Note that $E_1$
and $E_2$ are not necessary disjoint. The components of
$\redgr_{u,D}$ that do not contain $s$ and $t$ are called
\emph{cyclic components}. When $D = \emptyset$, we simply refer to
$\redgr_{u,D}$ as the \emph{reduction graph of $u$}.

Thus the reduction graph is a `superposition' of two graphs on the
same set of vertices $V$: one graph with edges from $E_1$ (reality
edges), and one graph with edges from $E_2$ (desire edges). The
following example should make the notion of reduction graph more
clear.

\def\pijlr#1{\ar@/^0.3pc/[r]^{\delta_{#1}}}
\def\pijll#1{\ar@/^0.3pc/[l]^{\bar \delta_{#1}}}
\def\pijlru#1{\ar@/^0.25pc/[ru]^{\delta_{#1}}}
\def\pijlld#1{\ar@/^0.25pc/[ld]^{\bar \delta_{#1}}}
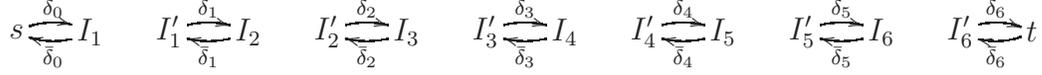
\begin{figure}
%
$$ \xymatrix @=16pt{
s           \pijlr{0} & \RGVertL{1} \pijll{0} &
\RGVertR{1} \pijlr{1} & \RGVertL{2} \pijll{1} &
\RGVertR{2} \pijlr{2} & \RGVertL{3} \pijll{2} &
\RGVertR{3} \pijlr{3} & \RGVertL{4} \pijll{3} &
\RGVertR{4} \pijlr{4} & \RGVertL{5} \pijll{4} &
\RGVertR{5} \pijlr{5} & \RGVertL{6} \pijll{5} &
\RGVertR{6} \pijlr{6} & t           \pijll{6} &
}$$
%
%
\caption{The part of the reduction graph of the legal string $u$
with respect to $D$ as defined in Example~\ref{reduction_graph_ex1}
which involves only reality edges (the vertex labels are omitted).}
\label{red_graph_ex1_1}
\end{figure}

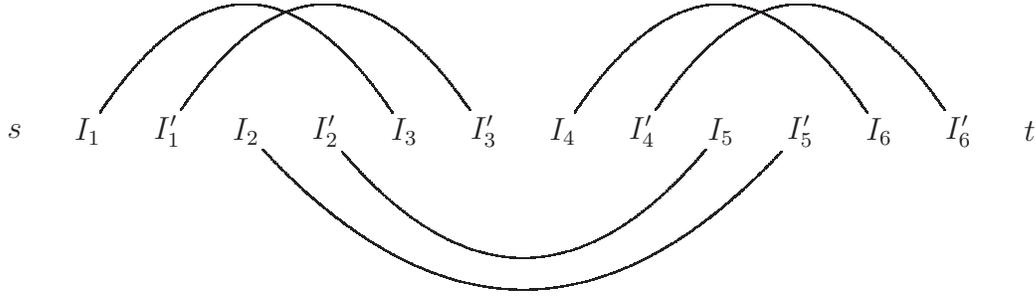
\begin{figure}
%
$$
\xymatrix @=16pt{
\\
\\
s &
\RGVertL{1} \ar@{-}@/^4pc/[rrrr]    & \RGVertR{1}
\ar@{-}@/^4pc/[rrrr] &
\RGVertL{2} \ar@{-}@/_5pc/[rrrrrrr] & \RGVertR{2}
\ar@{-}@/_4pc/[rrrrr] &
\RGVertL{3}                         & \RGVertR{3} &
\RGVertL{4} \ar@{-}@/^4pc/[rrrr]    & \RGVertR{4}
\ar@{-}@/^4pc/[rrrr] &
\RGVertL{5}                         & \RGVertR{5} &
\RGVertL{6}                         & \RGVertR{6} &
t \\
\\
\\
}$$
%
%
\caption{The part of the reduction graph of the legal string $u$
with respect to $D$ as defined in Example~\ref{reduction_graph_ex1},
where only desire edges are shown (the vertex labels are omitted).
Crossing edges correspond to positive pointers.}
\label{red_graph_ex1_2}
\end{figure}

\begin{figure}
%
$$
\xymatrix @=16pt{
\\
\\
s \pijlr{0} &
\RGVertL{1} \pijll{0} \ar@{-}@/^4pc/[rrrr] & \RGVertR{1} \pijlr{1}
\ar@{-}@/^4pc/[rrrr] &
\RGVertL{2} \pijll{1} \ar@{-}@/_5pc/[rrrrrrr] & \RGVertR{2}
\pijlr{2} \ar@{-}@/_4pc/[rrrrr] &
\RGVertL{3} \pijll{2} & \RGVertR{3} \pijlr{3} &
\RGVertL{4} \pijll{3} \ar@{-}@/^4pc/[rrrr] & \RGVertR{4} \pijlr{4}
\ar@{-}@/^4pc/[rrrr] &
\RGVertL{5} \pijll{4} & \RGVertR{5} \pijlr{5} &
\RGVertL{6} \pijll{5} & \RGVertR{6} \pijlr{6} & t \pijll{6} \\
\\
\\
}$$
%
%
\caption{The reduction graph $\redgr_{u,D}$ as defined in
Example~\ref{reduction_graph_ex1} (the vertex labels are omitted).}
\label{red_graph_ex1_3}
\end{figure}
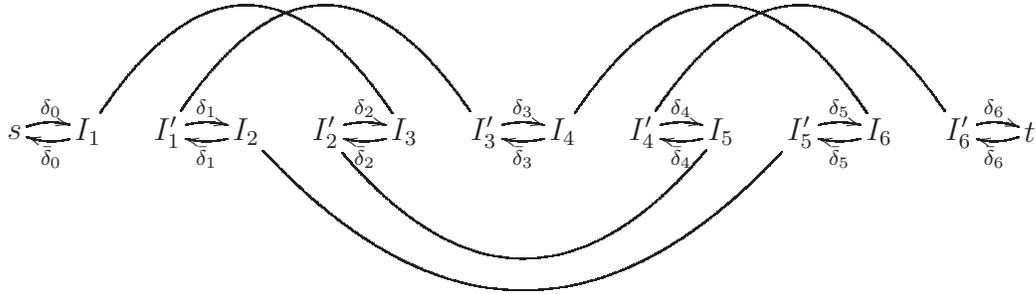

\def\pijlr#1{\ar@/^/[r]^{\delta_{#1}}}
\def\pijll#1{\ar@/^/[l]^{\bar \delta_{#1}}}
\def\pijlir#1{\ar@/^/[r]^{\bar \delta_{#1}}}
\def\pijlil#1{\ar@/^/[l]^{\delta_{#1}}}
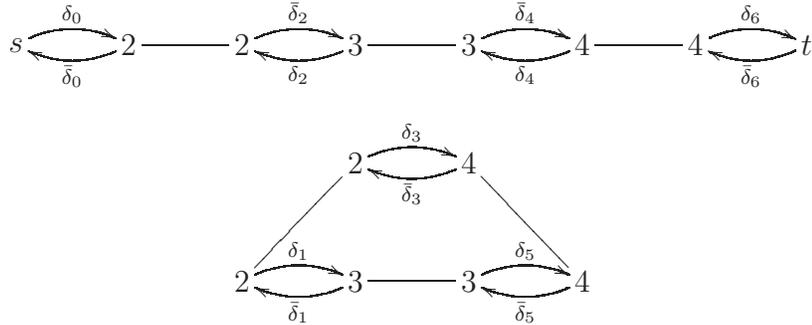
\begin{figure}
$$\xymatrix @=33pt{ s \pijlr{0} &
2 \pijll{0} \ar@{-}[r] & 2 \pijlir{2} &
3 \pijlil{2} \ar@{-}[r] & 3 \pijlir{4} &
4 \pijlil{4} \ar@{-}[r] & 4 \pijlr{6} & t \pijll{6} \\
& & & 2 \ar@{-}[dl] \pijlr{3} & 4 \pijll{3} \ar@{-}[dr] \\
& & 2 \pijlr{1} & 3 \pijll{1} \ar@{-}[r] & 3 \pijlr{5} & 4 \pijll{5} \\
}$$
\caption{The reduction graph of Figure~\ref{red_graph_ex1_3} where
every vertex (except $s$ and $t$) is represented by its label.}
\label{red_graph_ex2}
\end{figure}

\begin{Example}
\label{reduction_graph_ex1} Let $u = 5 2 6 8 8 3 \bar 2 5 \bar 4 3 7
7 4 6$ be a legal string and $D = \{5,6,7,8\} \subseteq dom(u)$.
Thus, $\{2,3,4\} = dom(u) \backslash D$, and
$$
u = \delta_0 \; 2 \; \delta_1 \; 3 \; \delta_2 \; \bar 2 \;
\delta_3 \; \bar 4 \; \delta_4 \; 3 \; \delta_5 \; 4 \; \delta_6
$$
with $\delta_0 = 5$, $\delta_1 = 688$, $\delta_2 = \lambda$,
$\delta_3 = 5$, $\delta_4 = \lambda$, $\delta_5 = 77$ and $\delta_6
= 6$. Notice that $\delta_1, \delta_2, \ldots, \delta_6 \in
\Pi_{D}^*$. This example corresponds to the situation in
Figure~\ref{fig_motivation_red_graph1}.

The reduction graph $\redgr_{u,D}$ of $u$ with respect to $D$ is
given in Figure~\ref{red_graph_ex1_3}. It is the union of the graphs
in Figure~\ref{red_graph_ex1_1} and Figure~\ref{red_graph_ex1_2}.
Note that for every desire edge $e$, we represent both $e$ and $\bar
e$ by a single unlabelled, undirected edge. The graphs are drawn in
a form that closely relates to the linear ordering of $u$. The
desire edges that cross correspond to positive pointers, and the
desire edges that do not cross correspond to negative pointers.

Since the exact identity of the vertices in a reduction graph is not
essential for the problems considered in this paper (we need only to
know, modulo `bar', which pointer is represented by a given vertex),
in order to simplify the pictorial notation of reduction graphs we
will replace the vertices (except for $s$ and $t$) by their labels.
Figure~\ref{red_graph_ex2} gives $\redgr_{u,D}$ in this way. In this
figure we have reordered the vertices, making it transparent that
$\redgr_{u,D}$ has a single cyclic component (the figure illustrates
why the adjective `cyclic' was added).
\end{Example}

Note that a reduction graph is an undirected graph in the sense that
if $e \in E_1$ ($e \in E_2$, resp.) then also $\bar e \in E_1$
($\bar e \in E_2$, resp.). If we think of a reduction graph as an
undirected graph by considering edges $e$ and $\bar e$ as one
undirected edge, then both $s$ and $t$ are connected to exactly one
(undirected) edge, and every other vertex is connected to exactly
two (undirected) edges. As as corollary to Euler's theorem, a
reduction graph has exactly one component that has a linear
structure with $s$ and $t$ as endpoints and possibly one or more
components that have a cyclic structure (the cyclic components).
Thus, there is a unique alternating walk from $s$ to $t$ in every
reduction graph.

If a 2-edge coloured graph $G$ has a unique alternating walk from
$s$ to $t$, then this walk is called the \emph{reduct of $G$},
denoted by $red(G)$. We know now that if $\redgr_{u,D}$ is a
reduction graph of a legal string $u$ with respect to $D \subseteq
dom(u)$, then the reduct exists. It is then also called the
\emph{reduct of $u$ to $D$}, and denoted by $red(u,D)$. Since
$\redgr_{u,dom(u)}$ consists of the vertices $s$ and $t$ connected
by a (reality) edge labelled by $u$ (and by $\bar u$ in the reverse
direction), we have $red(u,dom(u)) = u$. Also, it is clear that if
2-edge coloured graphs $G_1$ and $G_2$ are isomorphic, then
$red(G_1) = red(G_2)$.

\begin{Example}
\label{reduction_graph_ex2} If we take $u$ and $D$ from
Example~\ref{reduction_graph_ex1}, then
$$
red(u,D) = \delta_0 \bar \delta_2 \bar \delta_4 \delta_6 = 56,
$$
which is easy to see in Figure~\ref{red_graph_ex2}.
\end{Example}

\section{Reduction Function}
\label{paragr_rgr}
\newcommand{\rf}{r \! f}
Before we can prove (in the next section) our main theorem on
reducibility, we need to define reduction functions. A reduction
function operates on reduction graphs. As we will see, these
functions simulate the effect (up to isomorphism) of each of the
three string pointer reduction rules on a reduction graph. For a
vertex label $p$, the $p$-reduction function merges edges that form
a walk `over' vertices labelled by $p$ and removes all vertices
labelled by $p$.
\begin{Definition}
For each vertex label $p$, we define the \emph{$p$-reduction
function} $\rf_p$, which constructs for every 2-edge coloured graph
$G = (V, E_1, E_2, f, \Psi, s, t)$, the 2-edge coloured graph
$$
\rf_p(G) = (V', (E_1 \backslash E_{rem}) \cup E_{add}, E_2
\backslash E_{rem}, f|V', \Psi, s, t),
$$
with
\begin{eqnarray*}
V' & = & \{s,t\} \cup \{ v \in V \backslash \{s,t\} \mid f(v) \not=
p \},
\\
E_{rem} & = & \{ e \in E_1 \cup E_2 \mid f(\iota(e)) = p \mbox{ or }
f(\tau(e)) = p\}, \mbox{and}
\\
E_{add} & = & \{ (\iota(\pi),\ell(\pi),\tau(\pi)) \mid \mbox{$\pi =
e_1 e_2 \cdots e_n$ with $n > 2$ is an alternating walk}
\\
& & \mbox{in $G$ with $f(\iota(\pi)) \not= p$, $f(\tau(\pi)) \not=
p$, and $f(\tau(e_i)) = p$ for $1 \leq i < n$}\}.
\end{eqnarray*}
\mbox{ }
\end{Definition}

\begin{figure}
$$\xymatrix @=33pt{ s \ar@/^/[r]^{\delta_{0} \bar \delta_{2}} &
3 \ar@/^/[l]^{\delta_{2} \bar \delta_{0}} \ar@{-}[r] & 3
\pijlir{4} &
4 \pijlil{4} \ar@{-}[r] & 4 \pijlr{6} & t \pijll{6} \\
& & 3 \ar@/^/[r]^{\bar \delta_{1} \delta_{3}} \ar@{-}[d] & 4 \ar@/^/[l]^{\bar \delta_{3} \delta_{1}} \ar@{-}[d] \\
& & 3 \pijlr{5} & 4 \pijll{5} \\
}$$ \caption{The reduction graph obtained when applying $\rf_2$ to
the reduction graph of Figure~\ref{red_graph_ex2}.}
\label{red_graph_ex3}
\end{figure}
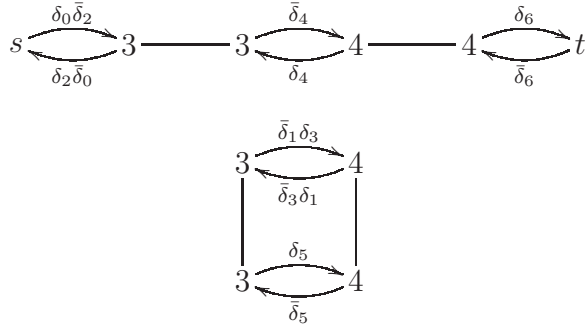

\begin{Example}
If we take the reduction graph $\redgr_{u,D}$ from
Example~\ref{reduction_graph_ex1}, cf. Figure~\ref{red_graph_ex2},
then $\rf_2(\redgr_{u,D})$ is given in Figure~\ref{red_graph_ex3}.
\end{Example}

It is easy to see that the following property holds for each
reduction graph $\redgr_{u,D}$ and all $p \in dom(u) \backslash D$:
$$
red( \redgr_{u,D} ) = red( \rf_p(\redgr_{u,D}) ).
$$
Also, reduction functions commute under composition. Thus, if
moreover there is a $q \in dom(u) \backslash D$ such that $p \not=
q$, then
$$
(\rf_q \ \rf_p)(\redgr_{u,D}) = (\rf_p \ \rf_q)(\redgr_{u,D}).
$$
The main property of reduction functions is that they simulate the
effect (up to isomorphism) of each of the three string pointer
reduction rules on a reduction graph.

\begin{Theorem} \label{rule_red_theorem}
Let $u$ be a legal string, let $D \subseteq dom(u)$, and let
$\varphi$ be a reduction of $u$ such that $dom(\varphi) = \{p_1,
p_2, \ldots, p_n\} \subseteq dom(u) \backslash D$. Then
$$
(\rf_{p_n} \ \cdots \ \rf_{p_2} \ \rf_{p_1})(\redgr_{u,D}) \approx
\redgr_{\varphi(u),D},
$$
and $red( u, D ) = red( \varphi(u), D )$.
\end{Theorem}
\begin{Proof}
To prove the first statement, it suffices to prove the cases where
$\varphi = {\bf snr}_p$, $\varphi = {\bf spr}_p$ and $\varphi = {\bf
sdr}_{p,q}$ for $p, q \in \Pi_{dom(u) \backslash D}$.

We first prove the ${\bf snr}$ case. Assume ${\bf snr}_p$ is
applicable to $u$. We consider the general case
$$
u = u_1 q_1 \delta_1 pp \delta_2 q_2 u_2
$$
for some $\delta_1,\delta_2 \in \Pi^*_D$, $q_1,q_2 \in \Pi_{dom(u)
\backslash D}$ and $u_1,u_2 \in \Pi^*$. In the special case where
$q_1$ ($q_2$, resp.) does not exist, the vertex labelled by
$\pset{q_1}$ ($\pset{q_2}$, resp.) in the graphs below equals the
source vertex $s$ (target vertex $t$, resp.). We will first prove
that $\rf_{\pset p} (\redgr_{u,D}) = \redgr_{{\bf snr}_p(u),D}$.
Because $u = u_1 q_1 \delta_1 pp \delta_2 q_2 u_2$, the reduction
graph $\redgr_{u,D}$ is
$$
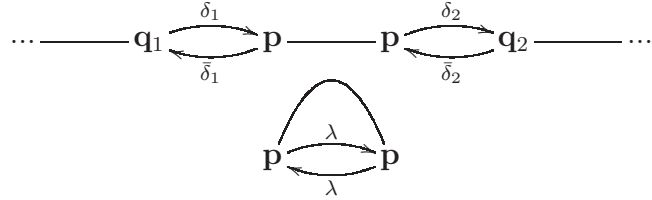
\xymatrix @=33pt{ ... \ar@{-}[r] & \pset{q_1} \pijlr{1} & \pset{p}
\pijll{1} \ar@{-}[r] & \pset{p} \pijlr{2} & \pset{q_2} \pijll{2} & ... \ar@{-}[l] \\
& & \pset{p} \ar@/^/[r]^{\lambda} \ar@{-}@/^2.5pc/[r] & \pset{p}
\ar@/^/[l]^{\lambda} }
$$
where we omitted the parts of the graph that remain the same after
applying $\rf_{\pset p}$. Now, the graph $\rf_{\pset p}
(\redgr_{u,D})$ is given below.
$$
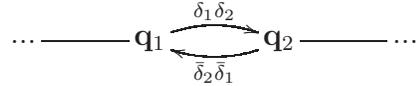
\xymatrix @=33pt{ ... \ar@{-}[r] & \pset{q_1} \ar@/^/[r]^{\delta_1
\delta_2} & \pset{q_2} \ar@/^/[l]^{\bar \delta_2 \bar \delta_1} &
... \ar@{-}[l] }
$$
This is clearly the reduction graph of ${\bf snr}_p(u) = u_1 q_1
\delta_1 \delta_2 q_2 u_2$ with respect to $D$. Thus, indeed
$\rf_{\pset p} (\redgr_{u,D}) \approx \redgr_{{\bf snr}_p(u),D}$.

We now prove the ${\bf spr}$ case. Assume ${\bf spr}_p$ is
applicable to $u$. We may distinguish three cases, which differ in
the number of elements of $\Pi_{dom(u) \backslash D}$ in between $p$
and $\bar p$ in $u$:
\begin{enumerate}
\item $u = u_1 q_1 \delta_1 p \delta_2 \bar p \delta_4 q_4 u_3$
\item $u = u_1 q_1 \delta_1 p \delta_2 q_2 \delta_3 \bar p
\delta_4 q_4 u_3$ \item $u = u_1 q_1 \delta_1 p \delta_2 q_2 u_2
q_3 \delta_3 \bar p \delta_4 q_4 u_3$
\end{enumerate}
for some $\delta_1,\ldots,\delta_4 \in \Pi^*_D$, $q_1,\ldots,q_4 \in
\Pi_{dom(u) \backslash D}$, and $u_1,u_2,u_3 \in \Pi^*$. Note that
we have assumed that $p$ is preceded and that $\bar p$ is followed
by an element from $\Pi_{dom(u) \backslash D}$. The special cases
where $q_1$ or $q_4$ do not exist, can be handled in the same way as
we did for the ${\bf snr}$ case (by setting them equal to $s$ and
$t$, resp.). In each of the three cases, one can prove that
$\rf_{\pset p} (\redgr_{u,D}) \approx \redgr_{{\bf spr}_p(u),D}$. We
will discuss it in detail only for the third case. The reduction
graph $\redgr_{u,D}$ is
$$
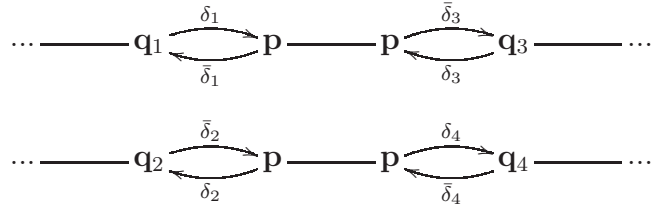
\xymatrix @=33pt{ ... \ar@{-}[r] & \pset{q_1} \pijlr{1} & \pset{p}
\pijll{1} \ar@{-}[r] & \pset{p} \pijlir{3} & \pset{q_3} \pijlil{3} & ... \ar@{-}[l] \\
... \ar@{-}[r] & \pset{q_2} \pijlir{2} & \pset{p} \pijlil{2}
\ar@{-}[r] & \pset{p} \pijlr{4} & \pset{q_4} \pijll{4} & ...
\ar@{-}[l] }
$$
where we again omitted the parts of the graph that remain the same
after applying $\rf_{\pset p}$. Now, the graph $\rf_{\pset p}
(\redgr_{u,D})$ is given below.
$$
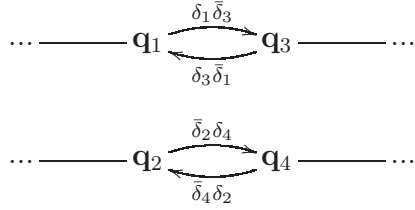
\xymatrix @=33pt{ ... \ar@{-}[r] & \pset{q_1} \ar@/^/[r]^{\delta_1
\bar \delta_3} & \pset{q_3} \ar@/^/[l]^{\delta_3 \bar \delta_1} &
... \ar@{-}[l] \\
... \ar@{-}[r] & \pset{q_2} \ar@/^/[r]^{\bar \delta_2 \delta_4} &
\pset{q_4} \ar@/^/[l]^{\bar \delta_4 \delta_2} & ... \ar@{-}[l] }
$$
This graph is clearly isomorphic to the reduction graph of
$$
{\bf spr}_p(u) = u_1 q_1 \delta_1 \bar \delta_3 \bar q_3 \bar u_2
\bar q_2 \bar \delta_2 \delta_4 q_4 u_3
$$
with respect to $D$. Thus, indeed $\rf_{\pset p} (\redgr_{u,D})
\approx \redgr_{{\bf spr}_p(u),D}$.

Finally, we prove the ${\bf sdr}$ case. Assume ${\bf sdr}_{p,q}$
is applicable to $u$. We only consider the general case (the other
cases are proved similarly):
$$
u = u_1 \; q_1 \delta_1 p \delta_2 q_2 \; u_2 \; q_3 \delta_3 q
\delta_4 q_4 \; u_3 \; q_5 \delta_5 p \delta_6 q_6 \; u_4 \; q_7
\delta_7 q \delta_8 q_8 \; u_5
$$
for some $\delta_1,\ldots,\delta_8 \in \Pi^*_D$, $q_1,\ldots,q_8 \in
\Pi_{dom(u) \backslash D}$, and $u_1,\ldots,u_5 \in \Pi^*$. The
reduction graph $\redgr_{u,D}$ is
$$
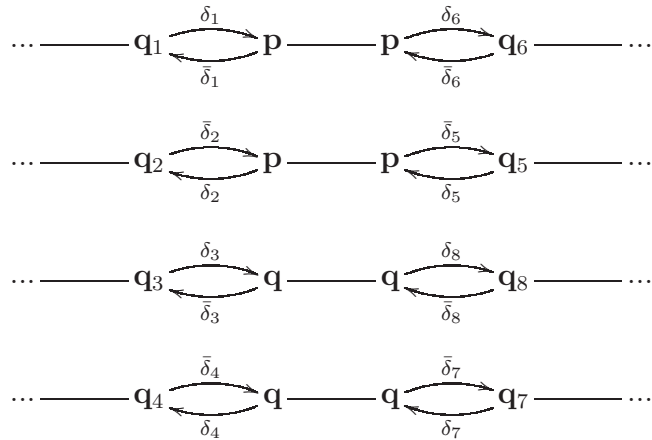
\xymatrix @=33pt{
... \ar@{-}[r] & \pset{q_1} \pijlr{1} & \pset{p} \pijll{1}
\ar@{-}[r] & \pset{p} \pijlr{6} & \pset{q_6} \pijll{6} & ... \ar@{-}[l] \\
... \ar@{-}[r] & \pset{q_2} \pijlir{2} & \pset{p} \pijlil{2}
\ar@{-}[r] & \pset{p} \pijlir{5} & \pset{q_5} \pijlil{5} & ... \ar@{-}[l] \\
... \ar@{-}[r] & \pset{q_3} \pijlr{3} & \pset{q} \pijll{3}
\ar@{-}[r] & \pset{q} \pijlr{8} & \pset{q_8} \pijll{8} & ... \ar@{-}[l] \\
... \ar@{-}[r] & \pset{q_4} \pijlir{4} & \pset{q} \pijlil{4}
\ar@{-}[r] & \pset{q} \pijlir{7} & \pset{q_7} \pijlil{7} & ... \ar@{-}[l] \\
}
$$
where we omitted the parts of the graph that remain the same after
applying $(\rf_{\pset q} \ \rf_{\pset p})$. Now, the graph
$\rf_{\pset q} ( \rf_{\pset p}(\redgr_{u,D}))$ is given below.
$$
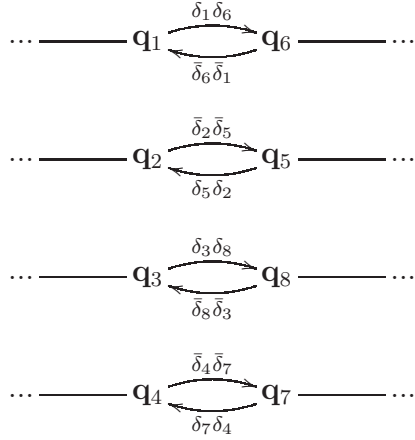
\xymatrix @=33pt{ ... \ar@{-}[r] & \pset{q_1} \ar@/^/[r]^{\delta_1
\delta_6} & \pset{q_6} \ar@/^/[l]^{\bar \delta_6 \bar \delta_1} &
... \ar@{-}[l] \\
... \ar@{-}[r] & \pset{q_2} \ar@/^/[r]^{\bar \delta_2 \bar
\delta_5} & \pset{q_5} \ar@/^/[l]^{\delta_5 \delta_2} & ...
\ar@{-}[l] \\
... \ar@{-}[r] & \pset{q_3} \ar@/^/[r]^{\delta_3 \delta_8} &
\pset{q_8} \ar@/^/[l]^{\bar \delta_8 \bar \delta_3} & ...
\ar@{-}[l] \\
... \ar@{-}[r] & \pset{q_4} \ar@/^/[r]^{\bar \delta_4 \bar
\delta_7} & \pset{q_7} \ar@/^/[l]^{\delta_7 \delta_4} & ...
\ar@{-}[l] \\
}
$$
This graph is clearly isomorphic to the reduction graph of
$$
{\bf sdr}_{p,q}(u) = u_1 q_1 \delta_1 \delta_6 q_6 u_4 q_7
\delta_7 \delta_4 q_4 u_3 q_5 \delta_5 \delta_2 q_2 u_2 q_3
\delta_3 \delta_8 q_8 u_5
$$
with respect to $D$. Thus, indeed $\rf_{\pset q} ( \rf_{\pset
p}(\redgr_{u,D})) \approx \redgr_{{\bf sdr}_{p,q}(u),D}$. This
proves the first statement.

Now, by the fact that the reduction function does not change the
reduct of the graph, and by the first statement, we have
$$
red( \redgr_{u,D} ) = red( (\rf_{p_1} \ \rf_{p_2} \ \cdots \
\rf_{p_n})(\redgr_{u,D}) ) = red( \redgr_{\varphi(u),D} ).
$$
Thus, $red( u, D ) = red( \varphi(u), D )$ and this proves the
second statement.
\end{Proof}

\section{Characterization of Reducibility}
\label{paragr_main_result} We are now ready to prove our main
theorem on reducibility. In Theorem~\ref{char1} we have shown that
if $u$ is reducible to $v$ in $S$, then $rem_{dom(v)}(u)$ is
successful in $S$. Here we strengthen this theorem into an iff
statement by additionally requiring that $v$ equals the reduct of
$u$ to $dom(v)$. The resulting characterization is independent of
the chosen set of reduction rules $S \subseteq \{Snr,Spr,Sdr\}$.

\begin{Theorem} \label{main_theorem} Let $u$ and $v$ be legal
strings, $D = dom(v) \subseteq dom(u)$ and $S \subseteq
\{Snr,Spr,Sdr\}$. Then $u$ is reducible to $v$ in $S$ iff
$rem_D(u)$ is successful in $S$ and $red(u,D) = v$.
\end{Theorem}
\begin{Proof}
Let $u$ be reducible to $v$ in $S$. Therefore, there is an
$S$-reduction $\varphi$ of $u$ such that $\varphi(u) = v$. Also,
$rem_D(u)$ is successful in $S$ by Theorem~\ref{char1}. By
Theorem~\ref{rule_red_theorem}, we have $red(u,D) =
red(\varphi(u),D)$. Now, $red(\varphi(u),D) = \varphi(u) = v$,
because $D = dom(\varphi(u))$.

To prove the reverse implication, let $rem_D(u)$ be successful in
$S$ and $red(u, D) = v$. We have to prove that $u$ is reducible to
$v$ in $S$. Clearly, there is a successful $S$-reduction $\varphi$
of $rem_D(u)$.

Assume that $\varphi$ is not applicable to $u$. Since $\varphi$ is
applicable to $rem_D(u)$, we know from Lemma~\ref{ops_appl} that
$\varphi = \varphi_2 \ {\bf snr}_p \ \varphi_1$ for some
$\varphi_1$, $\varphi_2$ and $p$, where $\varphi_1$ is applicable to
$u$ and ${\bf snr}_p$ is not applicable to $\varphi_1(u)$. Thus, $p
\delta p$ is a substring of $\varphi_1(u)$ with $\delta \in
\Pi_{D}^* \backslash \{ \lambda \}$. Therefore the following graph
$$
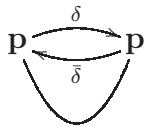
\xymatrix @=33pt{ \pset{p} \pijlr{} \ar@{-}@/_2.5pc/[r] & \pset{p}
\pijll{} }
$$
$$
$$
must be isomorphic to a cyclic component of the reduction graph
$\redgr_{\varphi_1(u),D}$ of $\varphi_1(u)$ with respect to $D$.
Because $v = red(u,D) = red(\varphi_1(u),D)$ is a legal string and
$dom(v) = D$, the labels of the reality edges of
$\redgr_{\varphi_1(u),D}$ belonging to cyclic components are empty.
This is a contradiction and therefore $\varphi$ is applicable to
$u$. Now, we have $\varphi(u) = red(\varphi(u), D) = red(u, D) = v$,
because $D = dom(\varphi(u))$. Thus, $u$ is reducible to $v$ in $S$.
\end{Proof}

Note that the proof of Theorem~\ref{main_theorem} even proves a
stronger fact. The $S$-reduction $\varphi$ of $u$ with $\varphi(u) =
v$ can be taken to be same as the (successful) $S$-reduction
$\varphi$ of $rem_D(u)$. The following corollary follows directly
from the previous theorem and the fact that every legal string is
successful in $\{Snr,Spr,Sdr\}$.
\begin{Corollary} \label{main_theorem_special}
Let $u$ and $v$ be legal strings and $D = dom(v) \subseteq
dom(u)$. Then $u$ is reducible to $v$ iff $red(u,D) = v$.
\end{Corollary}

The previous corollary shows that reducibility can be checked quite
efficiently. Since the reduction graph of a legal string $u$ has
$2|u| + 2$ vertices and $8|u| + 4$ edges (counting an undirected
desire edge as two (directed) edges), it takes only linear time
$O(|u|)$ to generate $\redgr_{u,\emptyset}$ using the adjacency
lists representation. Also, generating $\redgr_{u,D}$ for any $D
\subseteq dom(u)$ is of at most the same complexity as
$\redgr_{u,\emptyset}$. Now, since the walk from $s$ to $t$ does not
contain vertices more than once, it takes only linear time to
determine $red(u,D) = v$, and therefore, by the previous corollary,
it takes linear time to determine whether or not $u$ is reducible to
$v$.

The next corollary illustrates that the function of the reduct is
twofold: it does not only determine, given $u$ and $D \subseteq
dom(u)$, \emph{which} legal string is obtained by applying a
reduction $\varphi$ of $u$ with $dom(\varphi(u)) = D$, but also
\emph{whether or not} there is such a $\varphi$.
\begin{Corollary} \label{cor_applicability_reduct}
Let $u$ be a legal string and $D \subseteq dom(u)$. Then $u$ there
is a reduction $\varphi$ of $u$ with $dom(\varphi(u)) = D$ iff
$red(u,D)$ is legal and $dom(red(u,D)) = D$.
\end{Corollary}
\begin{Proof}
We first prove the forward implication. If we let $v = \varphi(u)$,
then $v$ is a legal string, $u$ is reducible to $v$, and $D =
dom(v)$. By Corollary~\ref{main_theorem_special}, $red(u,D) = v$ and
therefore $red(u,D)$ is legal and $dom(red(u,D)) = D$.

We now prove the reverse implication. If we let $v = red(u,D)$, then
$v$ is legal and $dom(v) = D$. By
Corollary~\ref{main_theorem_special}, $u$ is reducible to $v$.
\end{Proof}

\begin{Example}
Let $u$ and $D$ be as in Example~\ref{reduction_graph_ex1}. By
Example~\ref{reduction_graph_ex2}, $red(u,D) = 56$. Therefore by
Corollary~\ref{cor_applicability_reduct}, there is no reduction
$\varphi$ of $u$ with $dom(\varphi(u)) = D$. Thus, there is no
reduction $\varphi$ of $u$ with $dom(\varphi) = \{2,3,4\}$.
\end{Example}

\section{Cyclic Components} \label{paragr_cyclic_comp}
In this section we consider the cyclic components of the `full'
reduction graph $\redgr_{u,\emptyset}$ of a legal string $u$. We
show that if ${\bf snr}_p$ is applicable to $u$ for some pointer
$p$, then the number of cyclic components of $\redgr_{{\bf
snr}_p(u),\emptyset}$ is exactly one less than the number of cyclic
components of $\redgr_{u,\emptyset}$. On the other hand, if either
${\bf spr}_p$ or ${\bf sdr}_{p,q}$ is applicable to $u$ for some
pointer $p,q$, then the number of cyclic components remains the
same. Before we state this result
(Theorem~\ref{th_cyclic_components}), we will prepare for its proof
by studying some elementary connections between $u$ and the
structures in $\redgr_{u,\emptyset}$. Since all the edges of
$\redgr_{u,\emptyset}$ are labelled $\lambda$, we will omit the
labels of the edges in the figures.

Because desire edges in a reduction graph connect vertices that are
of the same label, for every label $\pset{p}$, there are exactly 0,
2 or 4 vertices labelled by $\pset{p}$ in every cyclic component of
a reduction graph. The following lemma establishes an additional
property of the number of vertices of a single label in a cyclic
component.
\begin{Lemma} \label{2vert_1label}
Let $u$ be a legal string, and let $P$ be a cyclic component in
$\redgr_{u,\emptyset}$. Let $p$ ($q$, resp.) be the first (last,
resp.) pointer (from left to right) in $u$ such that there is a
vertex in $P$ with label $\pset{p}$ ($\pset{q}$, resp.). Then there
are exactly two vertices of $P$ labelled by $\pset{p}$ and there are
exactly two vertices of $P$ labelled by $\pset{q}$.
\end{Lemma}
\begin{Proof}
Assume that all four vertices labelled by $\pset{p}$ are in $P$.
Then these vertices are $\RGVertL{i}$, $\RGVertR{i}$, $\RGVertL{j}$
and $\RGVertR{j}$ for some $i$ and $j$ with $i < j$. By the
definition of reduction graph, there is a reality edge from vertex
$\RGVertL{i}$ to vertex $\RGVertR{i-1}$. But by the definition of
$p$, vertex $\RGVertR{i-1}$ cannot belong to $P$, which is a
contradiction. Therefore, there are only two vertices labelled by
$\pset{p}$ in $P$. The second claim is proved analogously.
\end{Proof}

Note that in the previous lemma, $\pset{p}$ and $\pset{q}$ need not
be distinct. Note also that if all the vertices of a cyclic
component have the same label, than the cyclic component has exactly
two vertices.
\begin{Lemma} \label{lemma_cyclic_comp_2vert}
Let $u$ be a legal string, and let $p \in \Pi$. Then
$\redgr_{u,\emptyset}$ has a cyclic component consisting of exactly
two vertices, which are both labelled by $\pset{p}$ iff either $pp$
or $\bar p \bar p$ is a substring of $u$.
\end{Lemma}
\begin{Proof}
Let either $pp$ or $\bar p \bar p$ be a substring of $u$. Then
$$
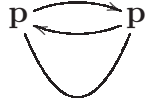
\xymatrix @=33pt{ \pset p \ar@/^/[r] \ar@{-}@/_2.5pc/[r] & \pset p
\ar@/^/[l] }
$$
$$
$$
is a cyclic component of $\redgr_{u,\emptyset}$ consisting of
exactly two vertices, both labelled by $\pset{p}$.

To prove the forward implication, let $\redgr_{u,\emptyset}$ have a
cyclic component $P$ consisting of exactly two vertices, both
labelled by $\pset{p}$. Clearly, every vertex of a cyclic component
has exactly one incoming and one outgoing edge in each colour.
Because there is a reality edge between the two vertices of $P$,
$\RGVertR{i}$ and $\RGVertL{i+1}$ are the vertices of $P$ for some
$i$. Now, since there is a desire edge $(\RGVertR{i},\RGVertL{i+1})$
in $P$, either $p$ or $\bar p$ occurs twice in $u$. As reality edges
in $\redgr_{u,\emptyset}$ connect adjacent pointers in $u$, either
$pp$ or $\bar p \bar p$ is a substring of $u$.
\end{Proof}

\begin{Lemma} \label{lemma_cyclic_comp_4vert}
Let $u$ be a legal string, let $p$ and $q$ be negative pointers
occurring in $u$. Then $\redgr_{u,\emptyset}$ has a cyclic component
consisting of exactly two vertices labelled by $\pset{p}$ and two
vertices labelled by $\pset{q}$ iff either $u = u_1 p q u_2 q p u_3$
or $u = u_1 q p u_2 p q u_3$ for some strings $u_1, u_2, u_3 \in
\Pi^*$.
\end{Lemma}
\begin{Proof}
Let either $u = u_1 p q u_2 q p u_3$ or $u = u_1 q p u_2 p q u_3$
for some strings $u_1, u_2, u_3 \in \Pi^*$. Then
$$
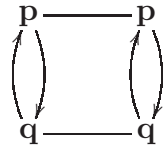
\xymatrix @=33pt{ \pset{p} \ar@/^/[d] \ar@{-}[r] & \pset{p} \ar@/^/[d] \\
\pset{q} \ar@/^/[u] \ar@{-}[r] & \pset{q} \ar@/^/[u]}
$$
is a cyclic component of $\redgr_{u,\emptyset}$ consisting of
exactly two vertices labelled by $\pset{p}$ and two vertices
labelled by $\pset{q}$.

To prove the forward implication, let $\redgr_{u,\emptyset}$ have a
cyclic component $P$ consisting of exactly two vertices labelled by
$\pset{p}$ and two vertices labelled by $\pset{q}$. Since each
cyclic component `is' a cycle of edges of alternating colour, and
since desire edges connect only vertices with the same label, the
component looks like the figure above. Since reality edges in
$\redgr_{u,\emptyset}$ connect adjacent pointers in $u$ and since
$p$ and $q$ are negative, either $u = u_1 pq u_2 qp u_3$ or $u = u_1
pq u_2 pq u_3$ with $u_i \in \Pi^*$ (with possibly $p$ and $q$
interchanged). Assume that $u = u_1 pq u_2 pq u_3$ (with possibly
$p$ and $q$ interchanged). Then there must be vertices $\RGVertR{i}$
and $\RGVertR{j}$ labelled by $\pset{p}$ with a desire edge
$(\RGVertR{i},\RGVertR{j})$ in $P$. But this is impossible since $p$
is negative. Consequently, $u = u_1 pq u_2 qp u_3$ (with possibly
$p$ and $q$ interchanged).
\end{Proof}

The following theorem states that only the string negative rules can
remove cyclic components. This is consistent with the fact that only
loop recombination introduces a new (cyclic) molecule, cf.
Figure~\ref{ld_op_fig}. Clearly, by the definition of reduction
function, a cyclic component is removed by simply removing its
vertices and edges and not by merging with another component.
\begin{Theorem} \label{th_cyclic_components}
Let $u$ be a legal string, let $N$ be the number of cyclic
components of $\redgr_{u,\emptyset}$, and let $p \in \Pi$ with
$\pset{p} \in dom(u)$.
\begin{itemize}
\item If ${\bf snr}_p$ is applicable to $u$, then the reduction
graph of ${\bf snr}_p(u)$ has exactly $N-1$ cyclic components.
\item If ${\bf spr}_p$ is applicable to $u$, then the reduction
graph of ${\bf spr}_p(u)$ has exactly $N$ cyclic components.
\end{itemize}
Now let $q \in \Pi$ with $\pset{q} \in dom(u)$ and $\pset{p} \not=
\pset{q}$.
\begin{itemize}
\item If ${\bf sdr}_{p,q}$ is applicable to $u$, then the
reduction graph of ${\bf sdr}_{p,q}(u)$ has exactly $N$ cyclic
components.
\end{itemize}
\end{Theorem}
\begin{Proof}
First note that by the definition of reduction function and
Theorem~\ref{rule_red_theorem} the number of cyclic components
cannot increase when applying reduction rules.

Let ${\bf snr}_p$ be applicable to $u$. By
Lemma~\ref{lemma_cyclic_comp_2vert}, $\redgr_{u,\emptyset}$ has a
cyclic component consisting of exactly two vertices, which are both
labelled by $\pset{p}$. It follows then from
Theorem~\ref{rule_red_theorem} that the reduction graph of ${\bf
snr}_p(u)$ has at most $N-1$ cyclic components. The other two
vertices labelled by $\pset{p}$ are connected by reality edges to
vertices that are not labelled by $\pset{p}$, and therefore this
component does not disappear. Hence, the reduction graph of ${\bf
snr}_p(u)$ has exactly $N-1$ cyclic components.

Let ${\bf spr}_p$ be applicable to $u$. Assume that the reduction
graph of ${\bf spr}_p(u)$ has less than $N$ cyclic components. Then
by Theorem~\ref{rule_red_theorem}, there exist a cyclic component
$P$ of $\redgr_{u,\emptyset}$ consisting of only vertices labelled
by $\pset{p}$. By Lemma~\ref{2vert_1label}, $P$ consists of only two
vertices. By Lemma~\ref{lemma_cyclic_comp_2vert}, either $pp$ or
$\bar p \bar p$ is a substring of $u$ and thus ${\bf spr}_p$ is not
applicable to $u$. This is a contradiction. Consequently, the
reduction graph of ${\bf spr}_p(u)$ has exactly $N$ cyclic
components.

Let ${\bf sdr}_{p,q}$ be applicable to $u$. Assume that the
reduction graph of ${\bf sdr}_{p,q}(u)$ has less than $N$ cyclic
components. Then there exist a cyclic component $P$ in
$\redgr_{u,\emptyset}$ consisting only of vertices labelled by
$\pset{p}$ and $\pset{q}$. Assume that all vertices of $P$ are
labelled by $\pset{p}$. Then, analogously to the previous case, we
deduce that either $pp$ or $\bar p \bar p$ is a substring of $u$.
Thus ${\bf sdr}_{p,q}$ is not applicable to $u$. This is a
contradiction. Similarly, $P$ cannot consist only of vertices
labelled by $\pset{q}$. Assume then that $P$ consists of vertices
that are labelled by both $\pset{p}$ and $\pset{q}$. By
Lemma~\ref{2vert_1label} and the fact that pointers $p$ and $q$
overlap, there are only two vertices labelled by $\pset{p}$ in $P$
and two vertices labelled by $\pset{q}$ in $P$. By
Lemma~\ref{lemma_cyclic_comp_4vert}, either $u = u_1 pq u_2 qp u_3$
or $u = u_1 qp u_2 pq u_3$ for some strings $u_1, u_2, u_3 \in
\Pi^*$. Thus ${\bf sdr}_{p,q}$ is not applicable to $u$. This is a
contradiction. Therefore, such a component $P$ cannot exist and so
the reduction graph of ${\bf sdr}_{p,q}(u)$ has exactly $N$ cyclic
components.
\end{Proof}

The previous theorem can be reformulated as follows, yielding a key
property of reduction graphs.
\begin{Theorem} \label{th_cyclic_components2}
Let $N$ be the number of cyclic components of the reduction graph of
legal string $u$. Then every successful reduction of $u$ has exactly
$N$ string negative rules.
\end{Theorem}

The Invariant Theorem~\cite{EPeo01-3} (and Chapter~12 in
\cite{GeneAssemblyBook}) shows that all successful reductions of a
\emph{realistic} string $u$ have the same number of string negative
rules. Therefore, Theorem~\ref{th_cyclic_components2} can be seen as
a generalization of this result, since it holds for \emph{legal}
strings in general. Indeed, the technical framework used in
\cite{EPeo01-3} is the MDS descriptor reduction system which is only
suited to model realistic strings.

Moreover, Theorem~\ref{th_cyclic_components2} shows that this number
$N$ is an elegant graph theoretical property of the reduction graph.
As a consequence, it can be efficiently obtained. Since it takes
$O(|u|)$ to generate $\redgr_{u,\emptyset}$, and again $O(|u|)$ to
determine the number of connected components of
$\redgr_{u,\emptyset}$, the previous theorem implies that it takes
only linear time to determine how many string negative rules are
needed to successfully reduce legal string $u$.
Theorem~\ref{th_cyclic_components2} will be used in the next
section, when we characterize successfulness in $S \subseteq
\{Spr,Sdr\}$.

\begin{Example}
Let $u = 2 3 \bar 2 \bar 4 3 4$ be a legal string. The reduction
graph of $u$ is depicted in Figure~\ref{red_graph_ex2}, where
$\delta_i = \lambda$ for all $i \in \{0,1,\ldots,6\}$. By
Theorem~\ref{th_cyclic_components2} every reduction of $u$ has
exactly one string negative rule. There are exactly four successful
reductions of $u$, these are ${\bf snr}_{2} \ {\bf spr}_{3} \ {\bf
spr}_{\bar 4}$, ${\bf snr}_{\bar 3} \ {\bf spr}_{2} \ {\bf
spr}_{\bar 4}$, ${\bf snr}_{\bar 3} \ {\bf spr}_{\bar 4} \ {\bf
spr}_{2}$ and ${\bf snr}_{4} \ {\bf spr}_{\bar 3} \ {\bf spr}_{2}$.
Notice that each of these reductions has exactly one string negative
rule.
\end{Example}

\begin{figure}
$$\xymatrix @=33pt{ s \ar@/^/[r] &
2 \ar@/^/[l] \ar@{-}[r] & 2 \ar@/^/[r] &
3 \ar@/^/[l] \ar@{-}[r] & 3 \ar@/^/[r] &
t \ar@/^/[l] \\
& 2 \ar@{-}@/_2.5pc/[r] \ar@/^/[r] & 2 \ar@/^/[l] & 3 \ar@/^/[r] \ar@{-}@/_2.5pc/[r] & 3 \ar@/^/[l] \\
\\
}$$
\caption{The reduction graph of $u = 2 2 3 3$.}
\label{red_graph_disj_ex_fig1}
\end{figure}
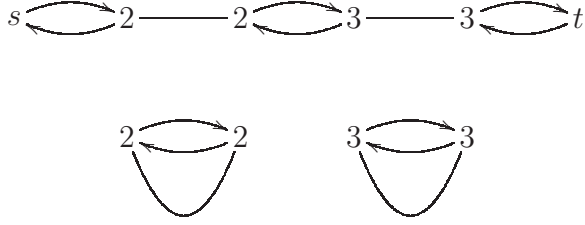

\begin{Remark} \label{remark_disjointcycle}
Results in \cite{SuccessfulnessChar_Original} (and Chapter~13 in
\cite{GeneAssemblyBook}) show that a successful reduction of a
realistic string $u$ has at least one string negative rule if the
string has a disjoint cycle. Clearly, the notions of disjoint cycle
and (cyclic) component are related. It is easy to verify that every
disjoint cycle of a string can be found as a connected component of
the reduction graph of the string, although that might be the linear
component. As an example, consider the realistic string $u =
\pi_3(M_1 M_2 M_3) = 2233$. This realistic string has three disjoint
cycles $\{22\}$, $\{33\}$, and $\{23,32\}$ corresponding to the
connected components of the reduction graph of $u$, see
Figure~\ref{red_graph_disj_ex_fig1}. This correspondence is not a
bijection for all legal strings, not even for realistic ones. E.g.,
realistic string $u = \pi_3(M_3 \bar M_1 M_2) = 3 \bar 2 2 3$ has
only a single disjoint cycle $\{33\}$ whereas its reduction graph
has two components, one linear and one cyclic. Hence, the number of
disjoint cycles cannot be used to characterize the number of string
negative rules present in every successful reduction of $u$.
\end{Remark}

It is easy to see that for legal string $u$ and $D \subseteq
dom(u)$, $\redgr_{rem_D(u), \emptyset}$ is isomorphic to
$\redgr_{u,D}$ modulo the labels of the edges. Now, we have the
following corollary to Theorems~\ref{th_cyclic_components2}.
\begin{Corollary}
Let $u$ be a legal string, $D \subseteq dom(u)$, and let $N$ be the
number of cyclic components of $\redgr_{u,D}$. Then every reduction
$\varphi$ of $u$ with $dom(\varphi(u)) = D$ has exactly $N$ string
negative rules.
\end{Corollary}
\begin{Proof}
Let $\varphi$ be a reduction of $u$ with $dom(\varphi(u)) = D$. Then
by Theorem~\ref{char1}, $\varphi$ is a successful reduction of
$rem_D(u)$. Since $\redgr_{u,D}$ is isomorphic to $\redgr_{rem_D(u),
\emptyset}$ modulo the labels of the edges, $\redgr_{rem_D(u),
\emptyset}$ has $N$ cyclic components. By
Theorem~\ref{th_cyclic_components2}, $\varphi$ has exactly $N$
string negative rules.
\end{Proof}

\section{Successfulness of Legal Strings}
\label{paragr_successfulness} In \cite{SuccessfulnessChar_Original}
(and Chapter~13 in \cite{GeneAssemblyBook}) an elementary
characterization of the \emph{realistic} strings that are successful
in any given $S \subseteq \{Snr,Spr,Sdr\}$ is presented. This is
helpful in applying Theorem~\ref{main_theorem}, where reducibility
of legal string $u$ into legal string $v$ is translated into
successfulness of $rem_D(u)$ with $D = dom(v)$. Unfortunately, even
when $u$ is a realistic string, $rem_D(u)$ for some $D \subseteq
dom(u)$ is not necessary a realistic string. For example, $u =
\pi_{5}(M_1 M_2 \bar{M}_3 M_4 M_5) = 2 2 3 \bar 4 \bar 3 4 5 5$ is
realistic, while $rem_{\{4\}}(u) = 2 2 3 \bar 3 5 5$ is not. As a
matter of fact, it can be shown that this legal string is not even
\emph{realizable}, that is, the legal string can not be transformed
into a realistic string by renaming pointers. Formally, legal string
$v$ is \emph{realizable} if there exists a homomorphism $h: \Pi
\rightarrow \Pi$ with $h(\bar p) = \overline{h(p)}$ for all $p \in
\Pi$ such that $h(v)$ is realistic. Thus, e.g., $2 2 3 \bar 3 4 4$
and $\bar 2 \bar 2 \bar 3 3 4 4$ are also not realistic.

In this section we generalize the results from
\cite{SuccessfulnessChar_Original}, and give a characterization of
the \emph{legal} strings that are successful in any given $S
\subseteq \{Snr,Spr,Sdr\}$. Theorems~\ref{th_snr_spr_char},
\ref{th_snr_sdr_char}, and \ref{th_snr_char} are the `legal
counterparts' of Theorems 8, 9, and 6 in
\cite{SuccessfulnessChar_Original}, respectively. These results are
independent of the results in the previous sections of this paper.
On the other hand, Theorems~\ref{th_spr_sdr_char},
\ref{th_sdr_char}, and \ref{th_spr_char} (the `legal counterparts'
of Theorems 14, 11, and 13 in \cite{SuccessfulnessChar_Original},
respectively) rely heavily on Theorem~\ref{th_cyclic_components2}.

\subsection{Trivial Generalizations and Known Results}

In the cases of $\{Snr, Spr\}$, $\{Snr, Sdr\}$, and $\{Snr, Spr,
Sdr\}$, the characterizations from
\cite{SuccessfulnessChar_Original} (and Chapter~13 in
\cite{GeneAssemblyBook}) and their proofs, although stated in terms
of realistic strings, are valid for legal strings in general. The
results are given below for completeness. First we restate Lemma~4
and Lemma~7 from \cite{SuccessfulnessChar_Original} respectively,
which will be used in our considerations below.

\begin{Lemma} \label{lemma_elem_strings}
Let $u = \alpha v \beta$ be a legal string such that $v$ is also a
legal string, and let $S \subseteq \{Snr, Spr, Sdr\}$. Then $u$ is
successful in $S$ iff both $v$ and $\alpha \beta$ are successful in
$S$.
\end{Lemma}

\begin{Lemma} \label{lemma_snr_spr_char}
Let $u$ be an elementary legal string. Then $u$ is successful in $\{
Snr, Spr \}$ iff either $u$ contains at least one positive pointer
or $u = pp$ for some $p \in \Pi$.
\end{Lemma}

The following result follows directly from
Lemma~\ref{lemma_elem_strings} and Lemma~\ref{lemma_snr_spr_char}.
It is the `legal version' of Theorem~8 in
\cite{SuccessfulnessChar_Original}, which can be taken almost
verbatim.
\begin{Theorem} \label{th_snr_spr_char}
Let $u$ be a legal string. Then $u$ is successful in $\{Snr, Spr\}$
iff for all legal substrings $v$ of $u$, if $v = v_1 u_1 v_2 \cdots
v_j u_j v_{j+1}$, where each $u_i$ is a legal substring, then $v_1
v_2 \cdots v_{j+1}$ either contains a positive pointer or is
successful in $\{Snr\}$.
\end{Theorem}

The previous theorem can be stated more elegantly in terms of
connected components of the overlap graph of $u$, see
\cite[p.141]{GeneAssemblyBook}. Note that characterization for case
$\{Snr, Spr\}$ refers to the case of $\{Snr\}$. The latter case does
differ from the realistic characterization in
\cite{SuccessfulnessChar_Original}, and is treated later.

\begin{Theorem} \label{th_snr_sdr_char}
Let $u$ be a legal string. Then $u$ is successful in $\{Snr,
Sdr\}$ iff all the pointers in $u$ are negative.
\end{Theorem}

We give now the legal version of Theorem~9.1 in
\cite{GeneAssemblyBook} --- it is a direct consequence of
Theorems~\ref{th_snr_spr_char} and \ref{th_snr_sdr_char}. Without
restrictions on the types of reduction rules used, every legal
string is successful, cf. the remark below the definition of the
reduction rules, in Section~\ref{paragr_sprs}.
\begin{Theorem}
Every legal string is successful in $\{Snr,Spr,Sdr\}$.
\end{Theorem}

\subsection{Non-Trivial Generalizations}

The following theorem is the legal counterpart of Theorem~6 in
\cite{SuccessfulnessChar_Original}. It turns out to be much less
restrictive than the original realistic version.
\begin{Theorem}
\label{th_snr_char} Let $u$ be a legal string. Then $u$ is
successful in $\{Snr\}$ iff $u$ consists of negative pointers only
and no two pointers overlap in $u$.
\end{Theorem}
\begin{Proof}
The condition from the statement of the lemma is obviously
necessary, because ${\bf snr}$ cannot resolve overlapping or
positive pointers. We will now prove that this condition is also
sufficient. If no two pointers overlap in $u$, then there must be a
substring $pp$ or $p \bar p$ of $u$ for some pointer $p$. If
moreover $u$ consists of negative pointers only, then $pp$ is a
substring of $u$. So ${\bf snr}_p$ is applicable to $u$. Now, again
no two pointers overlap in legal string ${\bf snr}_p(u)$, and ${\bf
snr}_p(u)$ consists of negative pointers only. By iteration of this
argument we conclude that $u$ is successful in $\{Snr\}$.
\end{Proof}
Observe that the $\{Snr\}$ case is referred to in the
characterization of $\{Snr, Spr\}$ in Theorem~\ref{th_snr_spr_char}.
With the above result we can rephrase the latter result as follows.

\begin{Corollary}
Let $u$ be a legal string. Then $u$ is successful in $\{Snr, Spr\}$
iff for all legal substrings $v$ of $u$, if $v = v_1 u_1 v_2 \cdots
v_j u_j v_{j+1}$, where each $u_i$ is a legal substring, then, if
$v_1 v_2 \cdots v_{j+1}$ consists of negative pointers only, they
are nonoverlapping.
\end{Corollary}

The following result follows directly from
Theorem~\ref{th_cyclic_components2}; a successful reduction without
string negative rules means that the reduction graph has a single
(linear) connected component.
\begin{Theorem} \label{th_spr_sdr_char}
Let $u$ be a legal string. Then $u$ is successful in $\{Spr,Sdr\}$
iff the reduction graph of $u$ has no cyclic component.
\end{Theorem}

Theorem~14 in \cite{SuccessfulnessChar_Original} is the realistic
predecessor of this result, but instead of cyclic components it uses
disjoint cycles, cf. Remark~\ref{remark_disjointcycle}. The latter
notion cannot be used in the general case, as, e.g., the legal
string $2 3 \bar 3 2 4 \bar 4$ has no disjoint cycle, but its
reduction graph has one cyclic component. Obviously, the only way to
reduce this string is to apply ${\bf spr}_3$ and ${\bf spr}_4$ (in
either order) and then to apply ${\bf snr}_2$. In particular, the
converse of Corollary 13.1 in \cite{GeneAssemblyBook} does not hold.

In the same way as Theorem~\ref{th_spr_sdr_char} relates to
Theorem~14 in \cite{SuccessfulnessChar_Original}, the following
theorem and lemma relate to Theorem~11 and Lemma~12 from
\cite{SuccessfulnessChar_Original}, respectively.

\begin{Theorem} \label{th_sdr_char}
Let $u$ be a legal string. Then $u$ is successful in $\{Sdr\}$ iff
$u$ consists of negative pointers only and $\redgr_{u,\emptyset}$
has no cyclic component.
\end{Theorem}
\begin{Proof}
The forward implication follows directly from
Theorem~\ref{th_cyclic_components2} and the fact that ${\bf sdr}$
cannot resolve positive pointers. To prove the reverse implication,
let $u$ consist of negative pointers only, and let the corresponding
reduction graph $\redgr_{u,\emptyset}$ have no cyclic component. By
Theorem~\ref{th_spr_sdr_char}, there is a successful
$\{Spr,Sdr\}$-reduction $\varphi$ of $u$. Since $u$ consists of
negative pointers only, $\varphi$ is a successful
$\{Sdr\}$-reduction of $u$ (as applications of string double rules
do not introduce positive pointers).
\end{Proof}

\begin{Lemma} \label{lemma_spr_char}
Let $u$ be an elementary legal string. Then $u$ is successful in
$\{Spr\}$ iff $u$ contains a positive pointer and
$\redgr_{u,\emptyset}$ has no cyclic component.
\end{Lemma}
\begin{Proof}
The forward implication follows directly from
Theorem~\ref{th_cyclic_components2}. To prove the reverse
implication, let $u$ contain a positive pointer and let
$\redgr_{u,\emptyset}$ have no cyclic component. By
Lemma~\ref{lemma_snr_spr_char}, there is a successful $\{Snr,
Spr\}$-reduction $\varphi$ of $u$. By
Theorem~\ref{th_cyclic_components2}, $\varphi$ is a
$\{Spr\}$-reduction of $u$.
\end{Proof}

The following result follows directly from
Lemmas~\ref{lemma_elem_strings} and \ref{lemma_spr_char} --- it
relates to Theorem~13 in \cite{SuccessfulnessChar_Original}.
\begin{Theorem} \label{th_spr_char}
Let $u$ be a legal string. Then $u$ is successful in $\{Spr\}$ iff
for all legal substrings $v$ of $u$, if $v = v_1 u_1 v_2 \cdots v_j
u_j v_{j+1}$, where each $u_i$ is a legal substring, then $v_1 v_2
\cdots v_{j+1}$ either is $\lambda$ or contains a positive pointer
and its reduction graph has no cyclic component.
\end{Theorem}

Similarly to Theorem~\ref{th_snr_spr_char}, the previous theorem can
be stated in terms of connected components of the overlap graph of
$u$.

Recall that for legal string $u$ and $D \subseteq dom(u)$,
$\redgr_{rem_D(u), \emptyset}$ is isomorphic to $\redgr_{u,D}$
modulo the labels of the edges. Then, by Theorems~\ref{main_theorem}
and \ref{th_spr_sdr_char}, we have the following corollary. In this
result it is especially apparent that both the linear component
\emph{and} the cyclic components of reduction graphs reveal crucial
properties concerning reducibility.
\begin{Corollary}
Let $u$ and $v$ be legal strings with $D = dom(v) \subseteq dom(u)$.
Then $u$ is reducible to $v$ in $\{Spr,Sdr\}$ iff $\redgr_{u,D}$ has
no cyclic component and $red(\redgr_{u,D}) = v$.
\end{Corollary}

\section{Discussion}
This paper introduces the concept of breakpoint graph (or reality
and desire diagram) into gene assembly models, through the notion of
reduction graph. The reduction graph provides surprisingly valuable
insights into the gene assembly process.
First, it allows one to characterize which gene patterns can occur
during the transformation of a given gene from its MIC form to its
MAC form. Formally, in the string pointer reduction system we
characterize whether a legal string $u$ is reducible to a legal
string $v$ for a given set of reduction rule types. The
characterization is independent from the chosen subset of the three
types of string pointer rules, and it allows us to determine whether
a legal string $u$ is reducible to a legal string $v$ in linear
time. This generalizes the characterization of successfulness in
\cite{SuccessfulnessChar_Original}, since the reduced string need
not be the empty string.
Secondly, the reduction graph allows one to determine the number of
loop recombination operations that are necessary in the
transformation of a given gene from its MIC form to its MAC form.
This result allows for a second generalization of the
characterization of successfulness, since we consider legal strings
instead of realistic strings.

Reduction graphs are defined for legal strings, the basic notion of
the string pointer reduction system that represents the genes.
Future research could focus on the possibility of defining a similar
notion for overlap graphs, which are used in the the graph pointer
reduction system --- a model (almost) equivalent to the string
pointer reduction system. This would allow results in this paper to
be carried over to the graph pointer reduction system.

\section*{Acknowledgments}
Grzegorz Rozenberg acknowledges support by NSF grant 0121422. The
authors are indebted to Tero Harju, Ion Petre, and the anonymous
referee for their valuable comments on this paper.

\bibliographystyle{plain}
\bibliography{gene_assembly}

\end{document}